\documentclass[useAMS,usenatbib]{mn2e}
\usepackage{graphicx}    
\usepackage{dcolumn}     
\usepackage{bm}          
\usepackage{amssymb,amsmath}  
\usepackage{color}
\usepackage{fixltx2e}

\title[Fast simulations for intensity mapping experiments]
      {Fast simulations for intensity mapping experiments}
\author[D. Alonso et al.]
  {David Alonso$^1$\thanks{E-mail:david.alonso@astro.ox.ac.uk}, Pedro G. Ferreira$^1$,
   Mario G. Santos$^{2,3,4}$ \\
   $^1$ Astrophysics, University of Oxford, DWB, Keble Road, Oxford OX1 3RH, UK\\
   $^2$ Department of Physics, University of Western Cape, Cape Town 7535, South Africa\\
   $^3$ SKA SA, 3rd Floor, The Park, Park Road, Pinelands, 7405, South Africa\\
   $^4$ CENTRA, Instituto Superior T\'ecnico, Universidade de Lisboa, Lisboa 1049-001, Portugal\\
  }

\begin{document}
  \date{\today}
  \pagerange{1--15} \pubyear{2014}
  \maketitle

  \begin{abstract}
    We present a code to generate mock observations of 21 cm intensity mapping experiments.
    The emphasis of the code is on reducing the computational cost of running a full-blown
    simulation, trading computational time for accuracy. The code can be used to generate
    independent realizations of the cosmological signal and foregrounds, which are
    necessary, for instance, in order to obtain realistic forecasts for future intensity
    mapping experiments. The code is able to reproduce the correct angular and radial
    clustering pattern for the cosmological signal, including redshift-space distortions,
    lightcone evolution and bias. Furthermore, it is possible to simulate a variety of
    foregrounds, including the potentially problematic polarized synchrotron emission.
  \end{abstract}


  \section{Introduction}\label{sec:intro}
    The distribution of perturbations in the matter density supplies a wealth of
    information on the late-time evolution of the Universe on cosmological scales. Until now,
    such studies have focused on optical redshift surveys \citep{Colless:2001gk,Jones:2004zy,
    Drinkwater:2009sd,2014MNRAS.441...24A}, which are able to trace the galaxy distribution to
    redshifts $z\lesssim1.5$ on relatively large scales, providing constraints on multiple
    observational probes, such as baryon acoustic oscillations (BAO) or redshift-space
    distortions (RSDs). One of the challenges of this type of observations is the large integration
    times that are necessary in order to obtain a good determination of the galaxy redshifts from
    their optical spectra. This limits the number density of sources with observed redshifts and
    restricts the maximum radial distance that can be reached. A possible alternative solution to
    this problem is to use only the photometry of each object to obtain a fast but inaccurate
    redshift measurement. Photometric redshift surveys \citep{Kaiser:2002zz,Tyson:2003kb,
    Abbott:2005bi} can thus observe a much higher number density of sources and reach larger
    redshifts (limited only by the sensitivity of each particular experiment), at the cost of
    losing almost all the relevant information in the radial direction ($\delta r_{\parallel}
    \sim\mathcal{O}(100\,{\rm Mpc}/h)$).
    
    A promising alternative has recently been proposed in the context of radio-astronomy.
    This approach, labelled {\it intensity mapping} \citep{Battye:2004re,2006ApJ...653..815M,
    Chang:2007xk,2008PhRvD..78b3529M,2008PhRvD..78j3511P,2008MNRAS.383..606W,2008MNRAS.383.1195W,
    2008PhRvL.100p1301L,2009astro2010S.234P,Bagla:2009jy,Seo:2009fq,2011ApJ...741...70L,
    2012A&A...540A.129A,2013MNRAS.434.1239B,2005ApJ...624L..65B,2014arXiv1405.1452B} is based on
    measuring the radio emission from different patches of the sky and different frequencies. Any
    pocket of neutral hydrogen will emit in the isolated 21cm line, corresponding to the hyperfine
    spin-flip transition of the 1s ground state. This is observed at a redshifted frequency
    $\nu_{\rm obs}=\nu_{21}/(1+z)$, where $\nu_{21}=1420.4\,{\rm MHz}$ is the rest-frame frequency
    of the 21cm line. Thus, by measuring the intensity of radio emission from different directions
    in the sky, it is in principle possible to trace the three-dimensional distribution of neutral
    hydrogen in the Universe. The aim of intensity mapping is therefore not to focus on measuring
    the flux of individual galaxies, but rather the combined emission arriving from relatively
    wide patches of the sky. Even though this technique misses any information related to the very
    small-scale perturbations, it is possible to make very fast 3D maps of the neutral hydrogen
    density on large scales, which are the most relevant ones for cosmological studies.
  
    Analyzing the data from any large-scale structure experiment requires extensive use of
    simulated catalogs to estimate statistical uncertainties, study potential sources
    of systematic errors, assess the model-independence of the results, etc. On the one hand,
    a large number of independent realizations is usually needed (e.g. $\mathcal{O}(100-1000)$
    for a good convergence of the covariance matrix in galaxy surveys \citep{2013MNRAS.428.1036M}).
    On the other hand, it is desirable to have mock catalogs which describe the observed field
    as accurately as possible. The computational cost of generating these mocks usually
    grows with the level of precision required by the experiment, and therefore a compromise must
    be reached between computational speed and accuracy. In particular, since 21cm intensity
    mapping can access large redshifts ($z\gtrsim2-3$), N-body simulations covering these volumes
    $(\sim500\,{\rm Gpc}^3)$ with a reasonable mass resolution $(\sim 10^{11}\,M_{\odot})$ can be an
    expensive solution. Along these lines, a lot of work has been done within the community of
    galaxy redshift surveys to develop alternative cheaper methods that can still reproduce the
    relevant physical observables \citep{2013MNRAS.435..743D,2013JCAP...06..036T,
    2014MNRAS.437.2594W}.
    
    On the other hand, one of the most important challenges facing HI intensity mapping is the
    presence of foregrounds (both galactic and extra-galactic) with amplitudes several orders of
    magnitude larger than the signal to be measured. The statistical properties, as well as
    the frequency dependence of these foregrounds differs significantly from those of the 
    signal, and therefore there is hope that they can be successfully subtracted
    \citep{2002ApJ...564..576D,2003MNRAS.346..871O,2005ApJ...625..575S,2006ApJ...648..767M,
    2006ApJ...650..529W,2008MNRAS.391..383G,2008MNRAS.389.1319J,2009MNRAS.398..401L,
    2009A&A...500..965B,2010A&A...522A..67B,
    2010MNRAS.409.1647J,2013ApJ...769..154M,2014MNRAS.441.3271W,2014ApJ...781...57S}.
    Nevertheless, this foreground subtraction is a potential source of systematic effects that
    could limit the observational power of intensity mapping for cosmology. Evaluating and
    modelling these systematics is therefore an essential step in the observational pipeline that
    requires the use of simulated realizations of these foregrounds.
    
    It has become the norm in the analysis of Cosmic Microwave Background (CMB) data to construct
    efficient simulations which can then be used to understand the analysis pipeline for any
    given experiment \citep{2013ApJS..208...19H,2013arXiv1303.5062P}. By including different
    foreground contaminants and instrumental systematic effects in the simulation, it is then
    possible, via Monte Carlo techniques, to accurately estimate the various biases that may
    enter the final result. It is in this spirit that we approach the problem of generating
    mock realizations of the 21cm emission by neutral  hydrogen (HI) after reionization ($z
    \lesssim4$), as well as its most relevant foregrounds. The method we present is similar
    to those used by several galaxy redshift surveys, and is based on generating a lognormal
    realization of the density field of neutral hydrogen. Through this method it is possible
    to implement different important effects (e.g.: the bias of HI with respect to the matter
    density, the lightcone evolution of the density field, redshift space distortions,
    frequency decorrelation in the foregrounds, etc.) at a very low computational cost.
    
    Section \ref{sec:theory} reviews the key theoretical results on which the method used to
    generate the cosmological signal is based, as well as the theoretical models used to validate
    the simulations. The actual method is explained in section \ref{sec:cosmosignal}. The models
    and procedures used to generate the foreground simulations are described in section
    \ref{sec:foregrounds}. We study the validity of these simulations in section
    \ref{sec:validation}. Finally, the main results of this study are summarized in section
    \ref{sec:discuss}.

  \section{Theoretical background}\label{sec:theory}
    \subsection{Linear perturbation theory}\label{ssec:lpt}
      The growth of structure via gravitational collapse is a non-linear process which, even
      in an expanding background, quickly departs significantly from the solutions of the
      linearized equations \citep{Peebles:1980}. However, linear perturbation theory is an
      extremely useful tool when studying structure formation, since it contains vital
      information which can be used to understand the late-time distribution of the matter
      density field. Furthermore, since, as we will see in the next section, the lognormal
      transformation is able to describe the distribution of the late-time non-linear density
      field in terms of the linear one, linear perturbation theory is, in our case, essential.

      In the Newtonian limit, the linearized equation that describes the evolution of the
      overdensity field $\delta$ is given by
      \begin{equation}\nonumber
        \ddot{\delta}+2H\dot{\delta}-\frac{3}{2}\frac{\Omega_M}{a^2}\delta=0,
      \end{equation}
      where the derivatives $(\dot{})$ are taken with respect to the cosmic time $t$,
      $\Omega_M$ is the non-relativistic matter parameter
      and $H\equiv \dot{a}/a$ is the expansion rate. Note that this equation is valid both in
      Fourier and real space, since the evolution of linear perturbations in the dust component
      is self-similar. The growing-mode solution to this equation is
      \begin{equation}
        \delta\propto g(a)\equiv\frac{5}{2}\Omega_M\,H_0^2\,H(a)\int_0^a\frac{da'}{[a'\,H(a')]^3},
      \end{equation}
      where $H_0$ is the local Hubble rate and we have used the normalization $g(a\ll1)\simeq a$.
      We will denote the growth factor normalized to its present value by $D(a)\equiv g(a)/g(1)$.
      Thus, at the linear level $\delta({\bf x},a)=\delta_0({\bf x})\,D(a)$, where ${\bf x}$ are
      comoving coordinates.

      Since we want to generate mock catalogs in redshift space, we are also interested in the
      peculiar velocity field, which causes redshift-space distortions (RSDs).
      For most interesting cases, the peculiar velocity field can be modelled as being irrotational
      to a good approximation (vector modes are predicted to decay rapidly), and therefore can be
      written as the gradient of a velocity potential ${\bf v}\simeq\nabla\varphi$. The velocity
      field is coupled to the overdensity field through the continuity equation, which reads
      \begin{equation}
        a\,f(a)\,H(a)\delta_{\bf k}-k^2\varphi_{\bf k}=0
      \end{equation}
      for a dust-like component, where $f(a)\equiv d\ln D/d\ln a$ is the growth rate. The redshift
      distortion is then given by the radial comoving peculiar velocity
      \begin{equation}
        \Delta z_{\rm RSD}=\frac{v_r}{c\,a}=\hat{\bf u}_r\cdot\frac{\nabla\varphi}{a}.
      \end{equation}

      For our purposes, the most useful result from linear perturbation theory for
      a dust component is that both the overdensity and the peculiar velocity fields
      can be self-similarly related to their present-day values
      \begin{equation}
        \delta({\bf x},z)=D(z)\,\delta_0({\bf x}),\hspace{6pt}
        {\bf v}({\bf x},z)=\frac{f(z)\,H(z)\,D(z)}{(1+z)\,H_0\,f_0}
        {\bf v}_0({\bf x}).
      \end{equation}
      These first-order perturbations are linearly related to the primordial
      perturbations, which seem to be Gaussianly distributed \citep{2013arXiv1303.5076P}.
      Therefore it is possible to produce realizations of the Fourier-space overdensity field by
      generating Gaussianly distributed random numbers with a variance given by the power spectrum
      \begin{equation}
        \langle\delta_{{\bf k}_1}\delta_{{\bf k}_2}\rangle\equiv
        \delta({\bf k}_1+{\bf k}_2)\,P({\bf k}).
      \end{equation}
      However, a Gaussian overdensity field cannot represent a realistic matter density, since, for
      one thing, values of $\delta<-1$ should have a probability of exactly 0.

    \subsection{Lognormal fields}\label{ssec:ln}
      Lognormal fields were first proposed and analyzed by \citet{Coles:1991if} as a possible model
      to describe the distribution of matter in the Universe.  A lognormal random field is defined
      in terms of a Gaussian random field through the local transformation
      \begin{equation}\label{eq:ln1}
        x_{\rm LN}=\exp x_{\rm G}.
      \end{equation}
      One of the nice properties of these fields is that, while the Gaussian variable $x_{\rm G}$
      is allowed to take any values in $(-\infty,+\infty)$, $x_{\rm LN}$ can only take positive
      values by construction. Furthermore, as discussed in \citet{Coles:1991if}, the density field
      evolved along Lagrangian trajectories according to the linear velocity field can be well
      described by a lognormal distribution, which justifies the use of lognormal fields from a
      physical point of view. In order to obtain a lognormal overdensity field with zero mean from
      a Gaussian field, the transformation (\ref{eq:ln1}) must be slightly varied, the correct
      equation being
      \begin{equation}\label{eq:ln2}
        1+\delta_{\rm LN}=\exp\left(\delta_G - \frac{\sigma_G^2}{2}\right),
      \end{equation}
      where $\sigma_G^2$ is the variance of the Gaussian overdensity field.
      
      Lognormal density fields have been used in the past by different collaborations to create
      fast mock galaxy catalogs using techniques similar to the ones described in section
      (\ref{sec:cosmosignal}) \citep{Cole:2005sx,Beutler:2011hx,2011MNRAS.415.2892B},
      and are, therefore, a well established tool. The simulation of the extragalactic radio sky
      by \citet{2008MNRAS.388.1335W} was also based on a similar method.
      
      Nevertheless, the lognormal transformation cannot be expected to describe the galaxy density
      correctly on arbitrarily small scales ($\lesssim 5\,{\rm Mpc}/h$,
      \citet{2010MNRAS.403..589K}), and therefore this technique is only useful for a limited
      number of analyses.

    \subsection{The intensity mapping power-spectrum}\label{sec:th_clustering}
      The intensity in an observed frequency bin $\delta\nu$ coming from the 21cm emission of an
      object at redshift $z$ with neutral hydrogen mass $M_{\rm HI}$, subtending a solid
      angle $\delta\Omega$ is given by \citep{2005MNRAS.360...27A}
      \begin{equation}\label{eq:M2T}
        I(\nu,\hat{\bf n})=\frac{3\,h_p\,A_{12}}{16\pi\,m_H}\frac{1}{((1+z)\,r(z))^2}
                \frac{M_{\rm HI}}{\delta\nu\,\delta\Omega}\nu_{21},
      \end{equation}
      where $A_{12}=2.876\times10^{-15}\,{\rm Hz}$ is the Einstein coefficient corresponding
      to the emission from the 21cm hyperfine transition, $h_p$ is Planck's constant and
      $m_{\rm H}=1.6733\times10^{-24}\,{\rm g}$ is the hydrogen atom mass. Here, $r(z)$ is the
      comoving curvature distance $r(z)=c\,{\rm sinn}(H_0\,\sqrt{|\Omega_k|}\,
      \chi(z)/c)/(H_0\,\sqrt{|\Omega_k|})$\footnote{We use the usual convention where ${\rm sinn}(x)$
      is $\sin(x)$, ${\rm sinh}(x)$ and $x$ for closed, open and flat universes respectively.}
      and $\chi(z)$ is the radial comoving distance
      \begin{equation}\label{eq:dzrel}
        \chi(z)=\int_0^z\frac{c\,dz'}{H(z')}.
      \end{equation}

      This intensity $I(\nu,\hat{\bf n})$ can be written in terms of a black-body
      temperature in the Rayleigh-Jeans approximation $T=I\,c^2/(2\,k_B\nu^2)$, where $k_B$
      is Boltzmann's constant. Using this we can estimate the mean brightness temperature
      coming from redshift $z$ and its fluctuations in terms of the neutral hydrogen density:
      \begin{equation}\label{eq:t2d}
        T_{21}(z,\hat{\bf n}) = (0.19055\, {\rm K})\,
        \frac{\Omega_{\rm b}\,h\,(1+z)^2\,x_{\rm HI}(z)}{\sqrt{\Omega_{\rm M}(1+z)^3+
        \Omega_{\Lambda}}}\,(1+\delta_{\rm HI}).
      \end{equation}
      Here $x_{\rm HI}(z)$ is the neutral hydrogen fraction (i.e. fraction of the total baryon
      density in HI) and $\delta_{\rm HI}$ is the HI overdensity field in redshift space (smoothed
      over the volume defined by $\delta\nu$ and $\delta\Omega$).
      
      \subsubsection{Angular clustering}
        For a given frequency band, the brightness temperature can be transformed into harmonic
        space
        \begin{equation}
          a_{lm}(\nu)=\int d\hat{\bf n}^2\,T(\nu,\hat{\bf n})\,Y_{lm}^*(\hat{\bf n}),
        \end{equation}
        where $Y_{lm}(\theta,\phi)$ are the spherical harmonic functions. The cross power-spectra
        between different frequency bands can then be related to the spectrum of perturbations of
        the matter density:
        \begin{align}\nonumber
          C_l(\nu_1,\nu_2)&\equiv\langle a_{lm}(\nu_1)\,a_{lm}^*(\nu_2)\rangle\\\label{eq:clth}
                      &=\frac{2}{\pi}\int_0^{\infty}dk\,k^2\,P_0(k)\,\omega_{l,1}(k)\,
                      \omega_{l,2}(k).
        \end{align}
        Here $P_0(k)$ is the $z=0$ matter power spectrum, and we have defined the window functions
        \begin{align}
          \omega_{l,i}(k)\equiv&\int_0^{\infty} dz\,\phi_i(z)\,D(z)\left[b(z)j_l(k\chi)-
                                                                         f(z)j_l''(k\chi)\right],
        \end{align}
        where $j_l$ is the $l$-th spherical Bessel function, $\phi_i$ is the selection function for
        the $i$-th frequency band (given in terms of redshift), $D(z)$ is the growth factor and
        $f(z)=d\ln D/d\ln a$ is the growth rate. Here we have used the Kaiser approximation
        \citep{Kaiser:1987qv} for RSDs and have assumed that the HI density is linearly biased with
        respect to the matter density ($\delta_{\rm HI}=b(z)\,\delta_{\rm M}$).
      
        For the results shown here, the selection functions $\phi_i(z)$ are just top-hat windows in
        the redshift range covered by a given frequency band. The effect of an instrumental beam or
        a finite pixel size can be included in the $C_l$'s by scaling them with the corresponding
        angular window function $C_l\rightarrow |b_l|^2\,C_l$, where $b_l$ are the Legendre
        coefficients of the real space window $b(\theta)$:
        \begin{equation}
          b_l\equiv 2\pi\int_{-1}^1L_l(x)\,b(x\equiv \cos\theta)\,dx.
        \end{equation}
        For top-hat and Gaussian beams, for instance, the respective window functions are
        \begin{align}
          &b^{\rm TH}_l=\frac{L_{l-1}(\cos\theta_{\rm TH})-L_{l+1}(\cos\theta_{\rm TH})}
                            {(2l+1)\,(1-\cos\theta_{\rm TH})},\\\,
          &b^{\rm G}_l=\exp\left(-\frac{l(l+1)\,\theta_{\rm G}^2}{2}\right)
        \end{align}
        
      \subsubsection{Radial clustering.}\label{sssec:rad_cluster}
        Formally, the only way to analyze the clustering pattern of the neutral hydrogen
        distribution along the line of sight is through the angular cross-correlation of
        different frequency channels. The main reason for this limitation is the fact that, while
        the matter distribution is expected to be isotropic, it is not homogeneous in the radial
        direction when observed on the lightcone: the growth of structure as well as the bias and
        the neutral hydrogen fraction, for example, evolve with $z$.
        
        In practice, however, it is possible to explore the radial distribution if one is willing
        to accept a few approximations. Within the flat-sky approximation and for an
        infinitesimally thin frequency bin, we can describe a given angular pixel as a circle of
        radius $R\simeq r(z)\,\Delta\theta$, where $\Delta\theta$ is the pixel angular resolution
        and $r(z)$ is the comoving distance to the redshift corresponding to the frequency $\nu$.
        Then we can assume that the temperature anisotropy in that pixel corresponds to the average
        temperature within this circle:
        \begin{equation}
          \delta_T(\nu,{\bf n}) \equiv \frac{\Delta T(\nu,{\bf n})}{\overline{T}(\nu)}= 
          \int\int_{C_R}\frac{d\,{\bf s}_{\perp}}{\pi\,R^2}
          \delta_{\rm HI}(r_{\parallel},{\bf r}_{\perp}-{\bf s}_{\perp}).
        \end{equation}
        Here we have made use of Eq. (\ref{eq:t2d}) and have defined the radial and angular
        coordinates $r_{\parallel}\equiv \chi(z)$, ${\bf r}_{\perp}=r(z)\,{\bf n}$. Note that all
        integrals with respect to transverse ($\perp$) vectors are 2-dimensional.
        
        Now let us assume that we take a finite frequency bin that is thick enough to study the
        radial clustering over the relevant scales but narrow enough so that we can assume that
        no significant evolution takes place within the corresponding range of redshifts. We can
        then study the 2-point statistics for pixels in the same angular location but corresponding
        to different frequencies within our bin. Let us first define the line-of-sight Fourier
        transform of $\delta_T$, which can be related to the Fourier modes of the three-dimensional
        overdensity field $\delta_{\rm HI}$:
        \begin{align}\nonumber
          \widetilde{\delta}_{\parallel}(k_{\parallel})&\equiv\int
          \frac{d r_{\parallel}}{\sqrt{2\pi}}\,
          \exp\left[i\,k_{\parallel}r_{\parallel}\right]\delta_T(r_{\parallel})\\
          &=\int\frac{d\,{\bf k}_{\perp}}{2\pi}\left(2\frac{J_1(k_{\perp}R)}{k_{\perp}R}\right)
          \widetilde{\delta}_{\rm HI}(k_{\parallel},{\bf k}_{\perp}).
        \end{align}
        Here $J_1(x)$ is the order-1 cylindrical Bessel function of the first kind and
        $\widetilde{\delta}_{\rm HI}(k_{\parallel},{\bf k}_{\perp})$ is the Fourier component of
        $\delta_{\rm HI}$ for the wave vector with LOS and transverse components $k_{\parallel}$
        and ${\bf k}_{\perp}$ respectively. We can then define the radial power spectrum 
        \begin{equation}
          \langle \widetilde{\delta}_{\parallel}(k_{\parallel})\,
                  \widetilde{\delta}^*_{\parallel}(k'_{\parallel})\rangle
          \equiv \delta^D(k_{\parallel}-k'_{\parallel})\,P_{\parallel}(k_{\parallel}),
        \end{equation}
        which is directly related to the three-dimensional spectrum of neutral hyrogen:
        \begin{equation}\label{eq:pkrad}
          P_{\parallel}(k_{\parallel})=\int_0^{\infty}\frac{dk_{\perp}\,k_{\perp}}{2\pi}\,
          \left[2\frac{J_1(k_{\perp}R)}{k_{\perp}R}\right]^2
          P_{\rm HI}(k_{\parallel},k_{\perp}).
        \end{equation}
        For the results shown in section \ref{sec:val_cosmo} we have used a simplified model
        for $P_{\rm HI}$, neglecting non-linearities and including RSDs through the Kaiser
        approximation:
        \begin{equation}
          P_{\rm HI}(k_{\parallel},k_{\perp})=D^2(z)
            \left(b(z) + f(z)\frac{k^2_{\parallel}}{k^2}\right)^2P_0(k)
        \end{equation}
        Note also that the effective finite width of the frequency shells used to calculate
        the power spectrum can be taken into account by scaling Eq. (\ref{eq:pkrad}) with
        the square of the window function
        \begin{equation}
          W(k_{\parallel}\delta \chi)=2\frac{\sin(k_{\parallel}\delta \chi/2)}
                                            {k_{\parallel}\delta \chi},
        \end{equation}
        where $\delta \chi=\frac{c\,(1+z)^2}{\nu_{21}\,H(z)}\delta\nu$ is the comoving scale
        corresponding to the shell width $\delta\nu$.

  \section{Simulating the cosmological signal}\label{sec:cosmosignal}
    Putting together the results described in sections \ref{ssec:lpt} and \ref{ssec:ln}, the
    method used by our code to generate fast mock HI catalogs is:
    \begin{enumerate}
      \item Consider a cubic box of comoving size $L$ and divide it into $N_{\rm grid}^3$
            cubical cells of size $l_c\equiv L/N_{\rm grid}$. This will determine
            the scales probed by the catalog: $2\pi/L \lesssim k \lesssim 2\pi/l_c$.
      \item Generate a realization of the Fourier-space Gaussian overdensity field at
            $z=0$ by producing Gaussian random numbers with variance
            \begin{equation}
              \sigma^2(k)\equiv\left(\frac{L}{2\pi}\right)^3P_0(k).
            \end{equation}
            This is done in a Fourier-space grid with ${\bf k}={\bf n}\,2\pi/L$ with
            $-N_{\rm grid}/2 \leq n_i \leq N_{\rm grid}/2$.
            
            At the same time, the $z=0$ velocity potential can be calculated from 
            the overdensity field as
            \begin{equation}
              \varphi_{\bf k}(z=0) = f_0\,H_0\,\frac{\delta_{\bf k}(z=0)}{k^2}.
            \end{equation}
      \item Transform these fields to configuration space using a Fast Fourier Transform,
            and calculate the radial velocity at each cell by projecting the gradient
            of the velocity potential along the line of sight (LOS). The direction of the
            LOS will depend on the position of the observer inside the box (since we want
            to produce full-sky catalogs we take this to be the center of the box). This will
            yield the Gaussian overdensity $\delta_G$ and radial velocity $v_r$ fields at $z=0$.
      \item Calculate the overdensity field and radial velocity in the lightcone by computing
            the redshift to each cell through the distance-redshift relation (Eq.
            (\ref{eq:dzrel})), and evolving the fields self-similarly to that redshift.
            
            At the same time we may perform the lognormal transformation on the Gaussian
            overdensity field. Thus, in a cell at ${\bf x}$ with redshift $z({\bf x})$, the
            overdensity and radial velocity are given by
            \begin{align}
              & 1+\delta_{\rm HI}({\bf x})=\exp\left[G(z)\delta_G({\bf x},z=0)-
                                            G^2(z)\,\sigma_G^2/2\right],\\
              & v_r({\bf x})=\frac{f(z)H(z)D(z)}{(1+z)\,f_0\,H_0}\,v_r({\bf x},z=0),
            \end{align}
            where $\sigma_G^2\equiv\langle\delta_G^2\rangle$ is the variance of the Gaussian
            overdensity at $z=0$ and the factor $G(z)\equiv D(z)\,b(z)$ accounts both for
            the growth of perturbations and for a possible linear galaxy bias $b$.
            
      \item We calculate the total HI mass stored in each cell through
            \begin{equation}\nonumber
              M_{\rm HI}=(2.775\times10^{11}\,M_{\sun})\,\left(\frac{l_c}{{\rm Mpc}/h}\right)^3
                         \frac{\Omega_{\rm b}\,x_{\rm HI}(z)}{h}\,(1+\delta_{\rm HI})
            \end{equation}

      \item We divide the box into spherical shells, which are in turn pixelized to yield maps
            of the 21cm brightness temperature at different frequency bands (corresponding to
            the width of the different radial shells). To each pixel we assign the temperature
            associated with the hydrogen mass enclosed within it (Eq. (\ref{eq:M2T})). This 
            implies interpolating between the cartesian mass grid and the spherical pixels. We
            do this interpolation through a Monte-Carlo integration: a number
            $N\sim10$ of points are randomly placed inside each cell, and a mass of
            $M_{\rm HI}/N$ is assigned to each of them. We then assign each point to a spherical
            shell and angular pixel according to its redshift and angular position. At this stage
            we implement RSDs by perturbing the cosmological redshift of each point with the
            redshift distortion corresponding to the radial velocity in the cell
            $\Delta z_{\rm RSD} = c\,(1+z)\,v_r({\bf x})$. Our code uses the {\tt HEALPix}
            pixelization scheme \citep{2005ApJ...622..759G}.
    \end{enumerate}
    
    As can be seen, the above recipe relies on two extra ingredients (besides the underlying
    cosmological model): we need a model for the evolution of the neutral hydrogen fraction
    $x_{\rm HI}$ and its bias with respect to the matter density $b(z)$.

  \section{Foregrounds}\label{sec:foregrounds}
    Probably the main challenge of 21cm observations for cosmology is the presence of galactic
    and extragalactic foregrounds with amplitudes several orders of magnitude larger than the
    cosmological signal - a situation which in many ways mimics that of the analysis of the CMB.
    However, the spectral smoothness of the foregrounds (or at least their clearly identifiable
    frequency dependence), should make it possible to subtract them.
    
    Five different types of foregrounds have been implemented in the present version of the code:
    unpolarized and polarized galactic synchrotron, galactic and extragalactic free-free emission
    and emission from extragalactic point radio sources. We have classified these in two
    categories, isotropic and anisotropic, according to their angular distribution on the sky, and
    different methods were used to simulate each of them. These methods are based on those used
    by other groups to simulate radio foregrounds \citep{2010MNRAS.409.1647J,2014ApJ...781...57S,
    2014arXiv1401.2095S}.
    
    We must note that the use of the term ``foregrounds'' when referring to these contaminants
    for intensity mapping can be a misnomer: some of the sources of these contaminants are in fact
    ``behind'' part of the HI signal (e.g. point sources can be found at very high redshifts).
    The use of this term, however, has become traditional and is widespread in the literature,
    and therefore we will use it here, trusting that it will not lead to confusion.
    
    \subsection{Isotropic foregrounds}\label{ssec:isotropic_foregrounds}
      \begin{table}
        \begin{center}
          \begin{tabular}{|c|c|c|c|c|}
            \hline
            Foreground              & A (mK$^2$) & $\beta$ & $\alpha$ & $\xi$ \\
            \hline
            Galactic synchrotron    & 700        & 2.4 & 2.80 & 4.0 \\
            Point sources           &  57        & 1.1 & 2.07 & 1.0 \\
            Galactic free-free      & 0.088      & 3.0 & 2.15 & 35  \\
            Extragalactic free-free & 0.014      & 1.0 & 2.10 & 35  \\
            \hline
          \end{tabular}
        \end{center}
        \caption{Foreground $C_l(\nu_1,\nu_2)$ model from \citet{2005ApJ...625..575S} for the pivot
                 values $l_{\rm ref}=1000$ and $\nu_{\rm ref}=130\,{\rm MHz}$.}
                 \label{tab:clmodel}
      \end{table}

      Assuming the foregrounds are Gaussianly distributed, any statistically isotropic emission can
      be entirely modelled in terms of the frequency-space angular power spectra:
      \begin{equation}\label{eq:cl_fg}
        \langle a_{lm}(\nu_1)\,a^*_{l'm'}(\nu_2)\rangle=\delta_{ll'}\,\delta_{mm'}\,
        C_l(\nu_1,\nu_2).
      \end{equation}
      We have simulated the emission from extragalactic point sources and free-free foregrounds
      according to this model\footnote{Obviously galactic free-free emission is not homogeneous.
      However, due to its exceptionally smooth frequency dependence and its subdominant amplitude,
      we do not believe a more sophisticated modelling is required at this stage.}.
      
      We have followed \citet{2005ApJ...625..575S} (SCK from here on) in modelling the power
      spectrum as
      \begin{equation}\label{eq:clsck}
        C_l(\nu_1,\nu_2)= A\,\left(\frac{l_{\rm ref}}{l}\right)^{\beta}\,
                             \left(\frac{\nu_{\rm ref}^2}{\nu_1\,\nu_2}\right)^{\alpha}
                             \exp\left(-\frac{\log^2(\nu_1/\nu_2)}{2\,\xi^2}\right).
      \end{equation}
      where $\xi$ is the frequency-space correlation length of the emission, which regulates its
      spectral smoothness (foregrounds with smaller $\xi$ will be less smooth in frequency, and
      will be therefore more challenging to subtract). The parameters used for the different
      foregrounds were taken from \citep{2005ApJ...625..575S} and are listed in Table
      \ref{tab:clmodel}.
      
      In order to include the frequency decorrelation we have:
      \begin{enumerate}
        \item Diagonalized the matrix $C_{ij}\equiv C_l(\nu_i,\nu_j)/[A\,(l_{\rm ref}/l)^{\beta}]$.
        \item Generated independent random Gaussian realizations in the diagonal basis with power
              spectrum
              \begin{equation}
                C_l^n=A\,\left(\frac{l_{\rm ref}}{l}\right)^{\beta}\,\lambda^n,
              \end{equation}
              where $\lambda^n$ is the $n-$th eigenvalue of $C_{ij}$.
        \item Rotated these realizations back to frequency-space.
      \end{enumerate}
      
      \subsubsection*{Note on point sources}
        \begin{figure}
          \centering
          \includegraphics[width=0.45\textwidth]{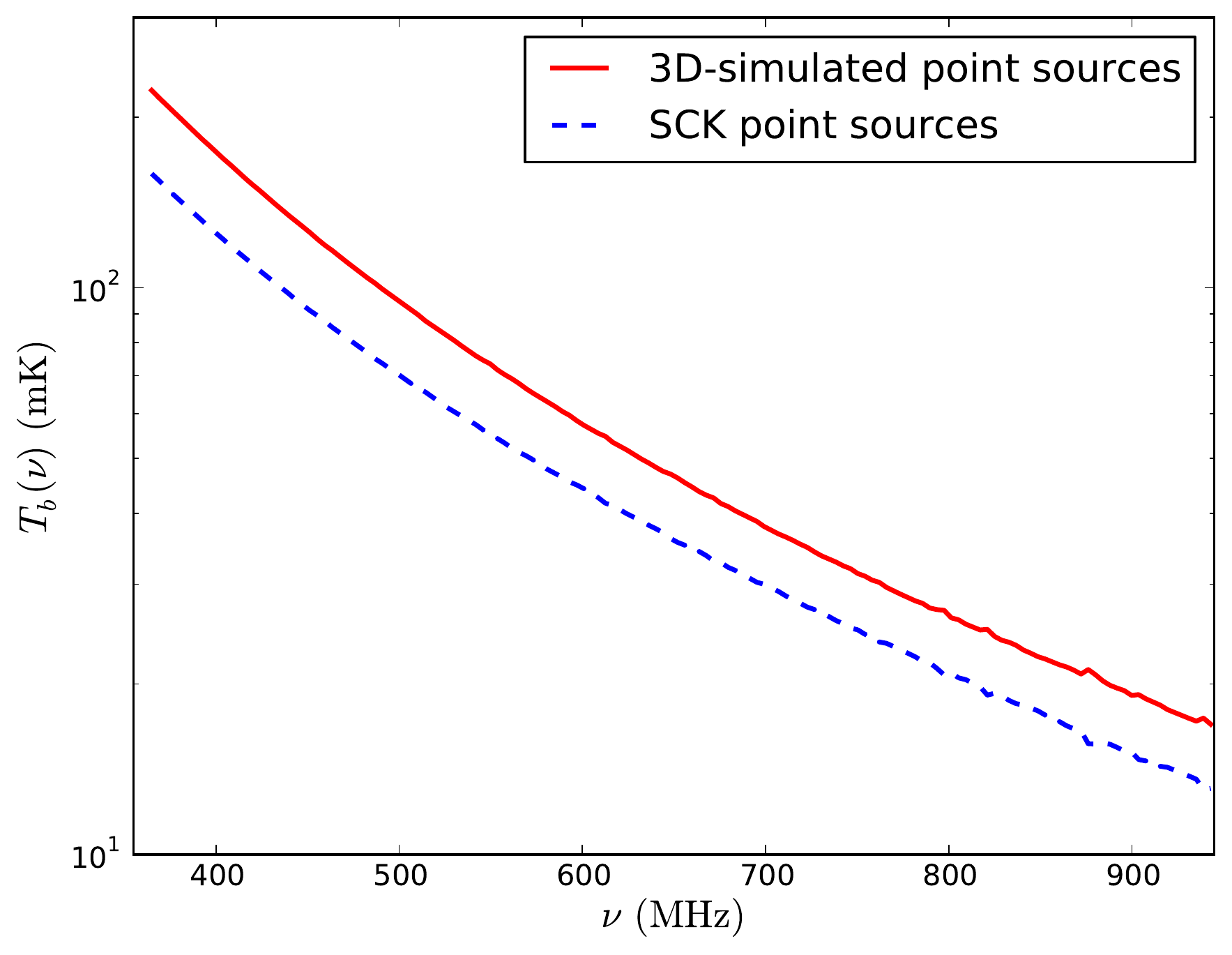}
          \caption{Brightness temperature as a function of frequency for a single
                   pixel of two of our simulations containing the cosmological
                   signal and two different realizations of point-source
                   foregrounds. The red solid line shows the result for point sources
                   simulated by sampling the density distribution that was used to
                   generate the cosmological signal, while the blue dashed line
                   correspond to a random realization of the SCK model.}
                   \label{fig:los_ps}
        \end{figure}
        \begin{figure}
          \centering
          \includegraphics[width=0.45\textwidth]{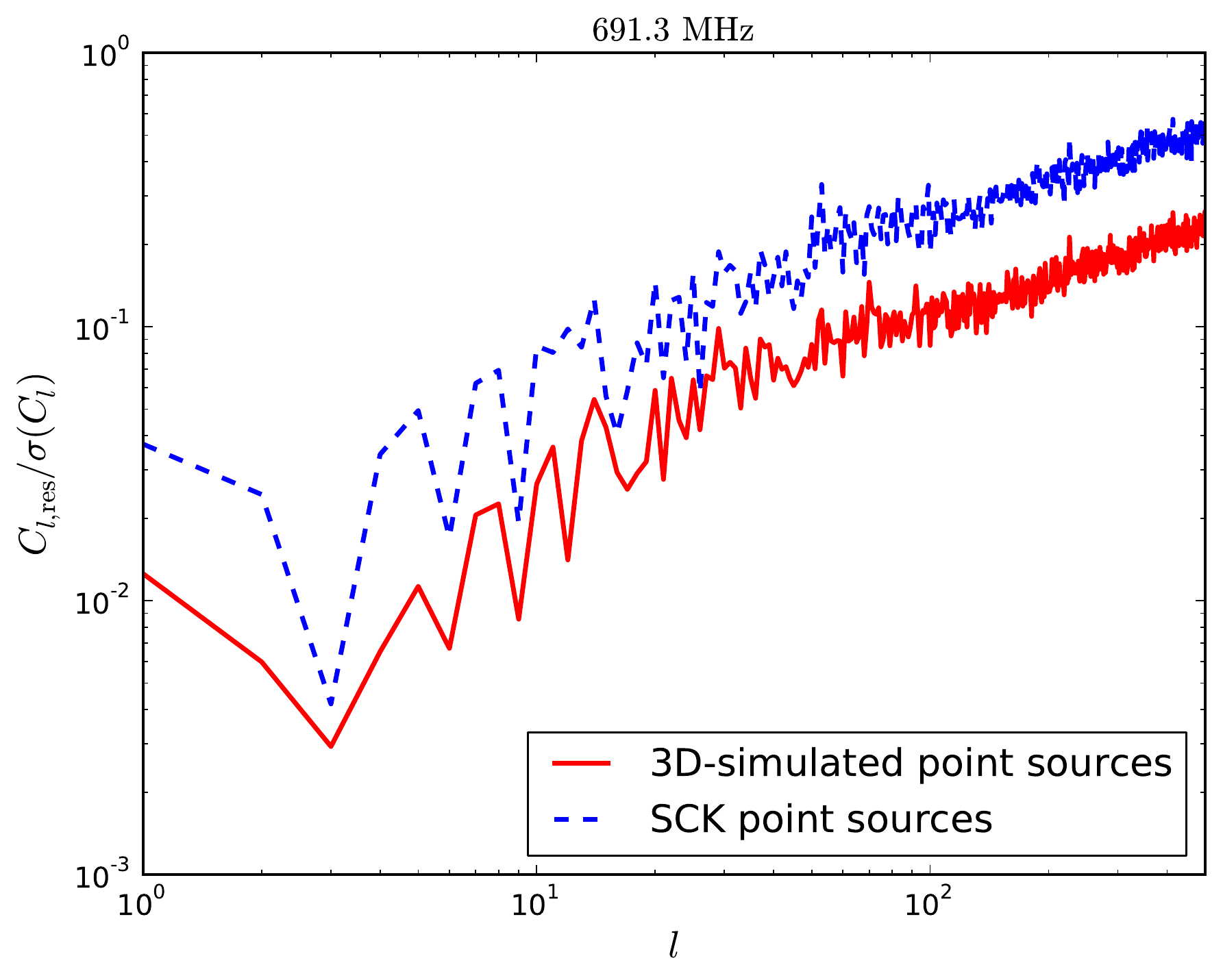}
          \caption{Ratio of the power spectrum of the foreground cleaning residuals to the
                   statistical errors in the cosmological power spectrum for a frequency bin
                   at $\bar{\nu}=691.3\,{\rm MHz}$. The red solid line shows the result for
                   maps containing point sources sampled from the density distribution, while
                   the blue dashed line corresponds to the SCK model for point sources.}
                   \label{fig:cl_ps}
        \end{figure}
        Extragalactic point sources are objects beyond our galaxy emitting in radio (e.g. AGNs).
        These objects should be clustered following the same matter distribution as the
        cosmological signal. Hence this foreground is potentially correlated with the
        cosmological signal if the redshift distribution of point sources overlaps with the
        redshift domain for which we aim to measure $\delta_{\rm HI}$ (which could be the case for
        intensity mapping). Therefore, in order to take this potential correlation into account,
        it would be desirable to be able to generate these point sources directly using the same
        matter distribution that was used for the cosmological signal. We have implemented this
        in a our code two main studies in mind.
        
        First, we would like to test the validity of approximating the emission from point sources
        as Gaussian realizations of the SCK model. To do so we compared this approach with a
        more realistic one: we Poisson-sampled the matter density field used for the cosmological
        signal with point sources following the luminosity function corresponding to star-forming
        galaxies that was used in \citet{2008MNRAS.388.1335W}. Each point source was
        given a luminosity and a random spectral index $\alpha$ following a normal distribution
        \begin{equation}
          P(\alpha)=\frac{1}{\sqrt{2\pi\,\alpha}}e^{-(\alpha-\alpha_0)^2/2\sigma_{\alpha}^2},
        \end{equation}
        with $\alpha_0=0.07$ \citep{2006ApJ...650..529W}. We verified that, even for large values
        of $\sigma_{\alpha}\sim 5-10$, the combined emission from point sources in the same angular
        pixel is extremely smooth and that in fact the frequency correlation length $\xi=1$ used
        above is a conservative lower bound.
        
        Secondly, since point sources are correlated with the cosmological signal, subtracting
        the former could potentially destroy (or degrade) the latter, an effect that our
        simulation, based on the SCK model, would not capture. We have attempted a small study of
        this effect here, although a more elaborate analysis is necessary to thoroughly address
        this potential issue.
        
        For a simulation of the cosmological signal (corresponding to the \textsc{Fast} parameters
        in table \ref{tab:sims}) we generated two simulations of the point-source  emission, one 
        based on the SCK model and another one produced by sampling the same density field and
        giving each point source a luminosity and spectral energy distribution (SED) as was done in
        \citet{2008MNRAS.388.1335W} for star-forming galaxies. We then produced two simplified
        observed sky maps by adding the cosmological signal and the two point-source simulations.
        For these simulations we also imposed a Gaussian beam with $\theta_{\rm FWHM}=0.3^o$ and
        uncorrelated Gaussian noise with $\sigma_T=0.05\,{\rm mK}$. Figure \ref{fig:los_ps} shows
        the brightness temperature in a single pixel as a function of frequency for both
        simulations.
        
        In order to study the potential loss of signal due to the subtraction of a foreground that
        is correlated with the signal, we cleaned the point-source foregrounds from these two
        simulations using the singular value decomposition method that was used in
        \citet{2013MNRAS.434L..46S}. For the SCK point-source maps the cleaning method converged
        after subtracting 6 foreground modes, while only 2 modes were necessary for the
        3D-simulated point sources. This is due to the fact that the SCK model assumes a very
        conservative frequency correlation length $\xi$ for point sources, while for the 3D
        simulation, every point source has the same SED. We then calculated the residual maps
        $T_{\rm clean}(\nu)-T_{\rm signal}(\nu)$ and computed their angular power spectra. We can
        measure the amount of signal loss by comparing these power spectra with the expected
        statistical errors in the cosmological signal:
        \begin{equation}
          \sigma(C_l)\simeq \sqrt{\frac{2}{2l+1}}\left(C_l+\frac{1}{\sigma_T^2}\right).
        \end{equation}
        We have plotted the ratio of these two quantities in figure \ref{fig:cl_ps}. As is
        shown, the residual $C_l$ is well below statistical uncertainties on all scales for both
        simulations, but a larger fraction of the signal is lost for the SCK foregrounds. This
        shows that the smoothness in frequency of the foregrounds is in fact more relevant than
        the potential correlation of point sources with the cosmological signal.

        The publicly available version of the code allows the user to simulate one population of
        point sources following a particular luminosity function, large-scale bias and spectral
        energy distribution. However we discourage the use of this feature, since the
        enormous number of point sources increases the computational time significantly and, as
        has been shown here, this method has no practical advantage over the empirical SCK model
        as far as foreground-cleaning studies are concerned.
        
      \subsubsection*{Line foregrounds}
        Another possible contaminant for HI intensity mapping could be due to the line emission
        of other sources (either galactic or extragalactic), and since this component is not
        spectrally smooth it might be impossible to subtract. However, one of the main advantages
        of the 21cm line is that it is very isolated in frequency, and therefore any source
        emitting in most of the potentially harmful lines would have to be at a very high
        redshift for it to contaminate the HI signal. Only a few molecular lines, such as the
        Hydroxyl radical (OH, $\sim1600-1700\,{\rm MHz}$) lie close to 21cm, and their signal
        is known to be very weak \citep{2011ApJ...740L..20G}. Therefore, unlike other lines,
        intensity mapping for HI should be virtually free of line confusion, and we have not
        attempted to include this contribution in our simulations.

    \subsection{Galactic synchrotron}\label{ssec:galactic_synchrotron}
      As can be seen in Table \ref{tab:clmodel}, galactic synchrotron is by far the largest
      contribution to the total emission in the frequencies of interest. For this reason it is very
      important to model it as realistically as possible.
      
      One of the key differences between galactic and extragalactic foregrounds is their angular
      structure: the intensity grows steeply towards the galactic plane, and everywhere the angular
      spectral tilt of galactic foregrounds is redder, making large angular structures more
      relevant. Subtracting a foreground with such a large structure could potentially erase or
      distort the cosmological signal on the largest angular scales, which could be a problem in
      order to measure large-scale observables, such as non-Gaussianity in the two-point
      clustering \citep{2013PhRvL.111q1302C}.
      
      Furthermore, linearly polarized radiation changes its polarization angle in as it traverses
      the galaxy due to Faraday rotation, an effect that is frequency-dependent and therefore not
      spectrally smooth. Hence, if part of the polarized foreground is leaked into the unpolarized
      part due to instrumental issues, it could be extremely difficult to subtract (e.g.
      \citet{2014PhRvD..89l3002D}).
      
      In this section we will first briefly describe the basics of galactic synchrotron emission
      and then explain the methods used in the simulations. These methods are largely based on
      those used by \citet{2014ApJ...781...57S,2014arXiv1401.2095S}.
      
      \subsubsection{Theoretical background} \label{sssec:synch_theory}
        A detailed description of the principles of galactic synchrotron emission can be found,
        for example, in \citep{1970ranp.book.....P,1986rpa..book.....R}.
        The galactic synchrotron radiation is caused by cosmic ray electrons interacting with
        the galactic magnetic field. The intensity of the emission coming from a volume element
        $dV=s^2ds\,\delta\Omega$ in a frequency interval $\delta\nu$ is given by the emission
        coefficient $j(s,\hat{\bf n},\nu)$. For the total and polarized synchrotron radiation, this
        is given by
        \begin{align}\nonumber
          &j_I(s,\hat{\bf n},\nu)=C_I\,\left(\frac{2\pi\,m_e\,c}{3e}\,
           \nu\right)^{\frac{1-p}{2}}n_{CR}\,B_{\perp}^{\frac{p+1}{2}},\\
          &j_P(s,\hat{\bf n},\nu)=C_P\,\left(\frac{2\pi\,m_e\,c}{3e}\,
           \nu\right)^{\frac{1-p}{2}}n_{CR}\,B_{\perp}^{\frac{p+1}{2}}\,e^{i\,2\phi_0},
        \end{align}
        Where $n_{CR}$ is the cosmic ray electron density, $B_{\perp}$ is the transverse
        galactic magnetic field, $\phi_0$ is the initial polarization angle and we are assuming a
        power-law energy distribution for the CR electrons $N(E)\propto E^{-p}$. The coefficients
        $C_{I,P}$ are given by
        \begin{align}\nonumber
          &C_I=\frac{\sqrt{3}e^3}{4\pi\,m_ec^2(p+1)}\Gamma\left(\frac{3p-1}{12}\right)\,
                                                   \Gamma\left(\frac{3p+19}{12}\right),\\
          &C_P=\Pi_0\,C_I,\hspace{12pt}\Pi_0=\frac{3\,p+3}{3\,p+7}
        \end{align}
        Assuming a spectral index $p\sim2.5$, the intrinsic polarized fraction is $\Pi_0\sim0.7$.
        Note that we have grouped the $Q$ and $U$ Stokes parameters into one single complex number
        labelled by $P$. The total intensity measured from the Earth is the line-of-sight integral
        of these emissivities.
        
        As the synchrotron photons traverse the magnetized interstellar medium, their polarization
        angle changes, an effect known as \emph{Faraday rotation} (see 
        \citet{1986rpa..book.....R} for further details). The observed polarization angle is thus
        not the same as the original one at the point of emission $\phi_0$. Both angles are related
        by $\phi=\phi_0+\psi(s,\hat{\bf n})\,\lambda^2$, where $\lambda=c/\nu$ and $\psi$ is the
        so-called Faraday rotation measure, given in terms of the longitudinal galactic magnetic
        field and the free electron density as
        \begin{equation}
          \psi(s,\hat{\bf n})=\frac{e^3}{2\pi\,(m_e\,c^2)^2}\int_0^sds'\,n_e(s',\hat{\bf n})\,
          B_{\parallel}(s',\hat{\bf n}).
        \end{equation}
        
        Thus, the polarized synchrotron intensity can be written as
        \begin{equation}\label{eq:synch_pol0}
          I_P(\nu,\hat{\bf n})=\Pi_0\int_0^{\infty}ds\,j_I(s,\hat{\bf n},\nu)
                         e^{2i\phi_0(s,\hat{\bf n})}\,
                         e^{i\psi(s,\hat{\bf n})x_{\nu}}, 
        \end{equation}
        where $x_{\nu}=2(c/\nu)^2$.
      
      \subsubsection{Unpolarized emission}\label{sssec:synch_unpol}
        The unpolarized synchrotron radiation should be spectrally smooth and, far from the
        galactic plane, we should be able to model it as an isotropic field, just like we did
        in section \ref{ssec:isotropic_foregrounds}. However, we would also like to include 
        the shape of the emission from the galactic plane, for which we have used the following
        recipe:
        \begin{enumerate}
          \item The Haslam map \citep{1982A&AS...47....1H}\footnote{We used the filtered and
                de-sourced map available at NASA's Legacy Archive for Microwave Background Data
                Analysis -{\tt http://lambda.gsfc.nasa.gov}.}
                contains the full-sky synchrotron emission at $\nu_H=408 {\rm MHz}$. This map is
                further filtered on angular scales $l\gtrsim200$ corresponding to its angular
                resolution ($\theta\sim1^o$). As a first approximation, we could simulate the
                emission in other frequencies by extrapolating the Haslam map with some spectral
                index. We have used the Planck Sky Model (PSM, \citet{2013A&A...553A..96D}) to
                generate an full sky map of the synchrotron spectral index $\beta(\hat{\bf n})$.
                With these two maps we calculate
                \begin{equation}
                  T_0(\nu,\hat{\bf n})=T_{\rm Haslam}(\hat{\bf n})
                  \left(\frac{\nu_H}{\nu}\right)^{\beta(\hat{\bf n})}.
                \end{equation}
	  \item Due to the poor resolution of the Haslam map, it would be desirable to
	        add structure on smaller scales as well as to introduce decorrelation between
	        different frequencies. We do this by generating a Gaussian realization of the $C_l$
	        model used in section \ref{ssec:isotropic_foregrounds}, with the parameters listed
	        in Table \ref{tab:clmodel}. These additional fluctuations should be constrained to
	        yield 0 on the scales already given by the Haslam map at $\nu=\nu_H$. This can be
	        guaranteed by constraining the eigenmode corresponding to the largest eigenvalue
	        of the matrix $C_{ij}$.
	        
	        Let us clarify this. The temperature fluctuation on the $i-$th frequency band is
	        given by
	        \begin{equation}
	          \delta T(\nu_i)=\sum_{n=1} B_i^n\,\delta\widetilde{T}_n,
	        \end{equation}
	        where $\hat{B}$ is the orthogonal matrix that diagonalizes $C_{ij}$ and
	        $\delta\widetilde{T}_n$ is the uncorrelated realization of the diagonal $C_l$'s
	        corresponding to the $n-$th eigenvalue of $C_{ij}$. We want $\delta T(\nu_H)=0$ on
	        the scales constrained by the Haslam map. Assuming that the highest eigenvalue
	        corresponds to $n=1$, we constrain the corresponding eigenmode on these scales to
	        be
	        \begin{equation}
	          \delta\widetilde{T}_1=-\sum_{n=2}\frac{B_{i_H}^n}{B_{i_H}^1}
	          \delta\widetilde{T}_n,
	        \end{equation}
	        where $i_H$ is the frequency bin corresponding to $\nu_H$.
	  \item The frequency-decorrelated unpolarized synchrotron emission including the galaxy
	        is then given by $T=T_0+\delta T$.
        \end{enumerate}

      \subsubsection{Polarized foregrounds}\label{sssec:synch_pol}
        The frequency-dependent Faraday rotation affecting polarized synchrotron emission makes
        any leakage of this foreground into the unpolarized signal a potentially dangerous
        contribution. Thus, in order to assess the feasibility of this subtraction or,
        alternatively the level of polarization leakage that can be allowed in any intensity
        mapping survey, it is necessary to have a correct description of this foreground.
      
        The existing observational data regarding polarized synchrotron emission is limited to
        isolated radio and microwave frequency bands \citep{2006A&A...448..411W,
        2008A&A...484..733T}, and the structure of the galactic magnetic field is poorly
        understood (e.g. \citet{2008ApJ...680..362H}). This situation will improve in the
        future (e.g. \citet{2009IAUS..259...89W}), but for the moment it is difficult to develop
        a reliable model of the polarized foregrounds in radio frequencies based only on
        observations. Alternatively, one can use existing models of the galactic magnetic field
        and cosmic ray and thermal electron densities in order to obtain more realistic
        predictions. This is the approach taken in {\tt Hammurabi} \citep{2009A&A...495..697W},
        a computer code that generates a 3D simulation of the Milky Way based on different models
        and then performs the line of sight integration to generate temperature maps at different
        frequencies. We have instead followed a different approach, based on the statistical
        properties of the synchrotron emission in the space of Faraday depths. As described in
        section \ref{sec:val_pol}, we have used {\tt Hammurabi} a posteriori to validate this
        model.
      
	For each line of sight (LOS) $\hat{\bf n}$, we can use the Faraday depth
	$\psi(s,\hat{\bf n})$ as LOS coordinate, instead of $s$, and rewrite Eq.
	(\ref{eq:synch_pol0}) as
	\begin{equation}\label{eq:synch_pol1}
	  I_P(\nu,\hat{\bf n})=\int d\psi\,k(\psi,\hat{\bf n},\nu)\,e^{i\psi x_{\nu}},
	\end{equation}
	where $k(\psi_0)=\int ds\, \delta(\psi(s)-\psi_0)\,j_I(s)\,e^{2i\phi_0(s)}$ is the
	collective emission from regions with Faraday depth $\psi$. As a first approximation we
	can assume that the spectral dependence of the emission is basically the same at all
	depths and can be factorized: $k(\psi,\hat{\bf n},\nu)=b(\nu,\hat{\bf n})\,
	k_0(\psi,\hat{\bf n})$
	
	\citet{2012A&A...542A..93O} used extragalactic point sources to measure the Faraday
	depth to the end of the Milky Way $\psi_{\infty}(\hat{\bf n})$. We can use their map
	to inspect the distribution of $\psi$ in different directions (see bottom left panel
	in Fig. \ref{fig:data}). Since in this map $\psi$ seems to take equally negative and
	positive values, we could model $\psi$ as being normally distributed around 0 with
	some variance $\sigma^2(\hat{\bf n})$. We can estimate this variance from the maps
	of $\psi_{\infty}$ by smoothing $\psi_{\infty}^2$ on a large angular scale (we have
	used $5^o$). Then, assuming that the collective emission at some $\psi$ is just
	proportional to the number of regions with that Faraday depth, we can model $k_0$ as
	\begin{equation}
	  k_0(\psi,\hat{\bf n})=B\,\exp\left[-\frac{1}{2}
	                        \left(\frac{\psi}{\sigma(\hat{\bf n})}\right)^2\right]\,
	                   \mu(\psi,\hat{\bf n})
	\end{equation}
	
	Lacking a better motivated model, we can assume that the field $\mu(\psi,\hat{\bf n})$
	has the same angular structure as the unpolarized emission and that it is correlated
	in Faraday space on scales smaller than some correlation length $\xi_{\psi}$
	\begin{equation}
	  \langle\mu_{lm}(\psi)\mu^*_{l'm'}(\psi')\rangle\propto\delta_{ll'}\delta_{mm'}
	  \left(\frac{l_{\rm ref}}{l}\right)^{\beta}e^{-\frac{1}{2}
	  \left[\frac{\psi-\psi'}{\xi_{\psi}}\right]^2},
	\end{equation}
	where the choice of a proportionality constant is degenerate with $B$. In order to
	diagonalize this covariance we can define the Fourier transform of $\mu(\psi)$
	\begin{equation}
	  \widetilde{\mu}(x)\equiv\int \frac{d\psi}{\sqrt{2\pi}}\, \mu(\psi)\,e^{i\,\psi x},
	\end{equation}
	which is uncorrelated for different values of $x$ with variance $\propto l^{-\beta}
	e^{-(\xi_{\psi}\,x)^2/2}$. It is then trivial to generate Gaussian realizations of the
	$\widetilde{\mu}(x)$'s.
	
	In terms of these functions, the integral in Eq. (\ref{eq:synch_pol1}) can be written as
	\begin{equation}\label{eq:synch_pol2}
	  I_P(\nu,\hat{\bf n})=B'\,b(\nu,\hat{\bf n})\int_{-\infty}^{\infty}dx\,
	  \widetilde{\mu}(x,\hat{\bf n})\, e^{-\frac{(x-x_{\nu})^2}{2\,\sigma^{-2}(\hat{\bf n})}}.
	\end{equation}
	Modelling the frequency dependence as
	$b(\nu,\hat{\bf n})=I_I(\nu,\hat{\bf n})/I_I(\nu_0,\hat{\bf n})$, we are
	left with two free parameters: the overall amplitude $B'$ and the correlation length in
	Faraday space $\xi_{\psi}$.
	
    \begin{figure*}
      \centering
      \includegraphics[width=0.90\textwidth]{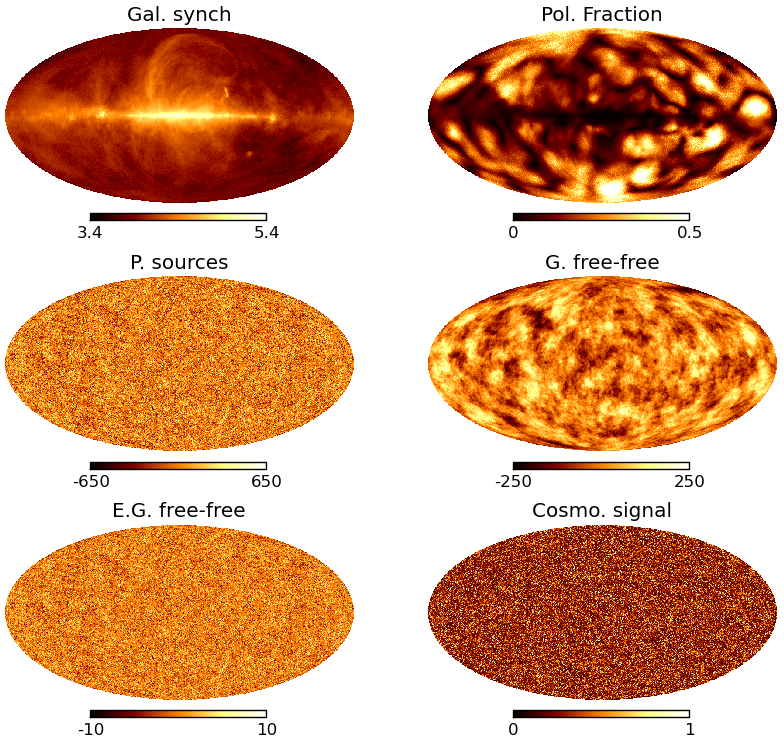}
      \caption{Full-sky maps of the different foregrounds and the cosmological signal for a
               frequency slice $\nu\sim565\,{\rm MHz}$. Temperatures are given in mK (with the
               top left plot showing $\log_{10}(T_{\rm synch})$), except in the case of the
               polarized fraction (upper right panel), which is dimensionless.}
               \label{fig:maps}
    \end{figure*}
	There are two observational facts that we would like our simulated catalogs to mimic:
	\begin{itemize} 
	  \item First, the synchrotron emission is more strongly depolarized closer to the galactic
	        plane. This is a sensible result: the magnetic field is stronger in that direction,
	        and there is a larger number of emitting regions adding up incoherently. The
	        ``width'' of a given line of sight in Faraday space is described in our simulation
	        by the value of $\sigma(\hat{\bf n})$, which we obtained from the map of
	        \citet{2012A&A...542A..93O}. Therefore this effect should be automatically
	        included.
	  \item At high frequencies ($x_{\nu}\rightarrow0$), the average depolarization at high
	        galactic latitudes is $\sqrt{|I_P|^2}/I_I\sim0.2-0.3$. We can use this observation
	        to fix the amplitude $B'$.
	\end{itemize}
	We are left with one free parameter, the correlation length in Faraday space $\xi_{\psi}$,
	which governs the frequency structure of the polarized emission. As described below
	(section \ref{sec:val_pol}), we have determined the value of this parameter by comparing
	the frequency dependence of our simulated maps to the one obtained using the more complex
	and realistic methods of {\tt Hammurabi}.
	
	To summarize, the method we have used to simulate the polarized synchrotron foreground is:
	\begin{enumerate}
	  \item Generate a Gaussian random realization of the $\mu(x,\hat{\bf n})$ with power
	        spectrum $C_l\propto l^{-\beta}\,e^{-x^2\xi_{\psi}^2/2}$ (arbitrary
	        normalization). The range and resolution in $x$ for these realizations will be
	        governed by the convergence of the integral in Eq. (\ref{eq:synch_pol2}) for
	        all values of $\sigma(\hat{\bf n})$.
	  \item For each line of sight $\hat{\bf n}$ and frequency $\nu$, we calculate the
	        integral in Eq. (\ref{eq:synch_pol2}) by summing over the realizations of $\mu$.
	        Note that each line of sight has its own ``Faraday width'' $\sigma(\hat{\bf n})$,
	        which will determine the relative depolarization of that LOS.
	  \item Fix the proportionality constant in Eq. (\ref{eq:synch_pol2}) by requiring the
	        average polarized fraction at large (microwave) frequencies and high
	        galactic latitudes to be $\sim0.2-0.3$, as measured by CMB experiments 
	        \citep{2007ApJ...665..355K}.
	\end{enumerate}
	
	We are mainly interested in the amount of polarized intensity that is leaked into the
	unpolarized part due to instrumental issues. The default version of our code implements
	this simply as a constant fraction of the Stokes parameter $Q$:
        \begin{equation}
          T_{\rm leak}(\nu,\hat{\bf n})=\epsilon_p\,T_{\rm syn}^Q(\nu,\hat{\bf n}).
        \end{equation}
        We would like to emphasize that this model is overly simple and not at allrealistic, since
        the leakage is defined with respect to telescope coordinates, and thus will correspond to
        different combinations of $Q$ and $U$ depending on the time of observation. Our aim here is
        to provide only an order-of-magnitude comparison between the polarization leakage and the
        cosmological signal (e.g. Figure \ref{fig:los_s}), but for any practical application a
        more precise parametrization would be necessary, using the full $Q$ and $U$ maps provided
        by the code.
        
  \section{Code validation}\label{sec:validation}
    \begin{table}
      \begin{center}
        \begin{tabular}{|l|c|c|}
          \hline
                                              & \textsc{Accurate} & \textsc{Fast} \\
          \hline
          $N_{\rm grid}$                          & 3072          &  2048         \\
          $L_{\rm box}\,({\rm Mpc}/h)$            & 8150          &  8900         \\
          $\nu$ interval (MHz)                    & $(405,945)$   &  $(355,945)$  \\
          $N_{\nu}$                               & 770           &  150          \\
          $z$ interval                            & $(0.5,2.5)$   &  $(0.5,3)$    \\
          $\delta r_{\parallel}\,\,({\rm Mpc}/h)$ & $\sim3.5$     &  $\sim20$     \\
          {\tt nside}                             & 512           &  512          \\
          Angular resolution (arcmin.)            & 6.7           &  6.7          \\
          Computational time                      & $\sim10$ h    & $\sim30$ min. \\
          Output size (GB, 1 field)               & 18            &  3.5          \\
          \hline
        \end{tabular}
      \end{center}
      \caption{Characteristics of the two types of simulations that were run for this work.
               The computational times correspond to real times using an 80-core shared-memory
               machine for the complete simulation (cosmological signal $+$ foregrounds).
               The output sizes correspond to one single field (e.g. the cosmological
               signal).} \label{tab:sims}
    \end{table}
    In order to validate the method, we have run several full simulations including the
    cosmological signal and foregrounds, and explored the qualitative behaviour of the
    different components. Figure \ref{fig:maps} shows the full-sky maps generated for
    the different fields in a frequency bin $\nu\in (563,567)\,{\rm MHz}$.
    
    We have generated two differend kinds of simulations with different spatial
    resolutions. The parameters for these are given in Table \ref{tab:sims}. The
    simulations labelled \textsc{Accurate} use a good spatial resolution ($l_c\simeq
    2.7\,{\rm Mpc}/h$) and narrow frequency bins ($\delta\nu\simeq 0.7\,{\rm MHz}$).
    They were used to compare their radial and angular power spectra with the
    theoretical predictions described in section \ref{sec:th_clustering}, as
    well as to evaluate the validity of our model for the polarized synchrotron
    foregrounds. The simulations labelled \textsc{Fast} were run with a coarser
    grid ($l_c\simeq 4.3\,{\rm Mpc}/h$) and wider frequency bins ($\delta\nu\simeq 4\,
    {\rm MHz}$). They represent a lower quality but faster version of the \textsc{Accurate}
    simulations and were used mainly to test the maximum speed that could be attained for a full
    valid simulation.
    
    We have not attempted to include any realistic instrumental effects in our simulations
    besides assuming a 1\% polarization leakage, since we would like the output to be
    applicable to any kind of experiment after a suitable post-processing. However, in order
    to show the lack of sensitivity of intensity mapping to small angular scales we have
    implemented an angular Gaussian beam of width $\theta_{\rm FWHM}=0.3^o$. At an intermediate
    redshift of 1.75 this corresponds to a comoving scale of about $7.6\,{\rm Mpc}/h$
    for the cosmological parameters below.
      
    \subsection{The cosmological signal}\label{sec:val_cosmo}
      \begin{figure}
        \centering
        \includegraphics[width=0.45\textwidth]{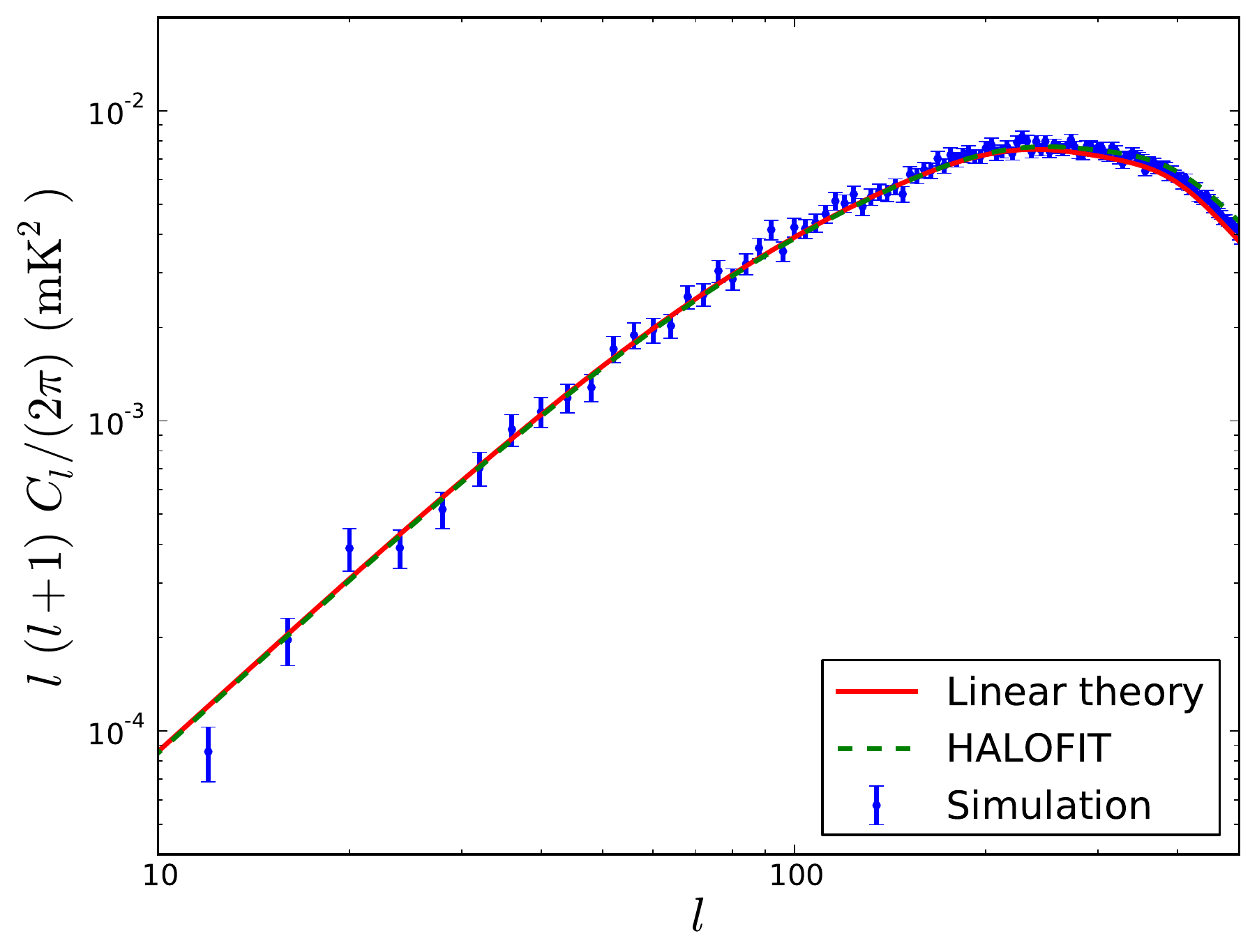}
        \caption{Angular power spectrum of the cosmological signal at $\nu\sim565\,{\rm MHz}$ as
                 measured in our simuation (blue dots with error bars) and according to linear
                 theory (solid red line).} \label{fig:cl_cosmo}
      \end{figure}
      \begin{figure}
        \centering
        \includegraphics[width=0.45\textwidth]{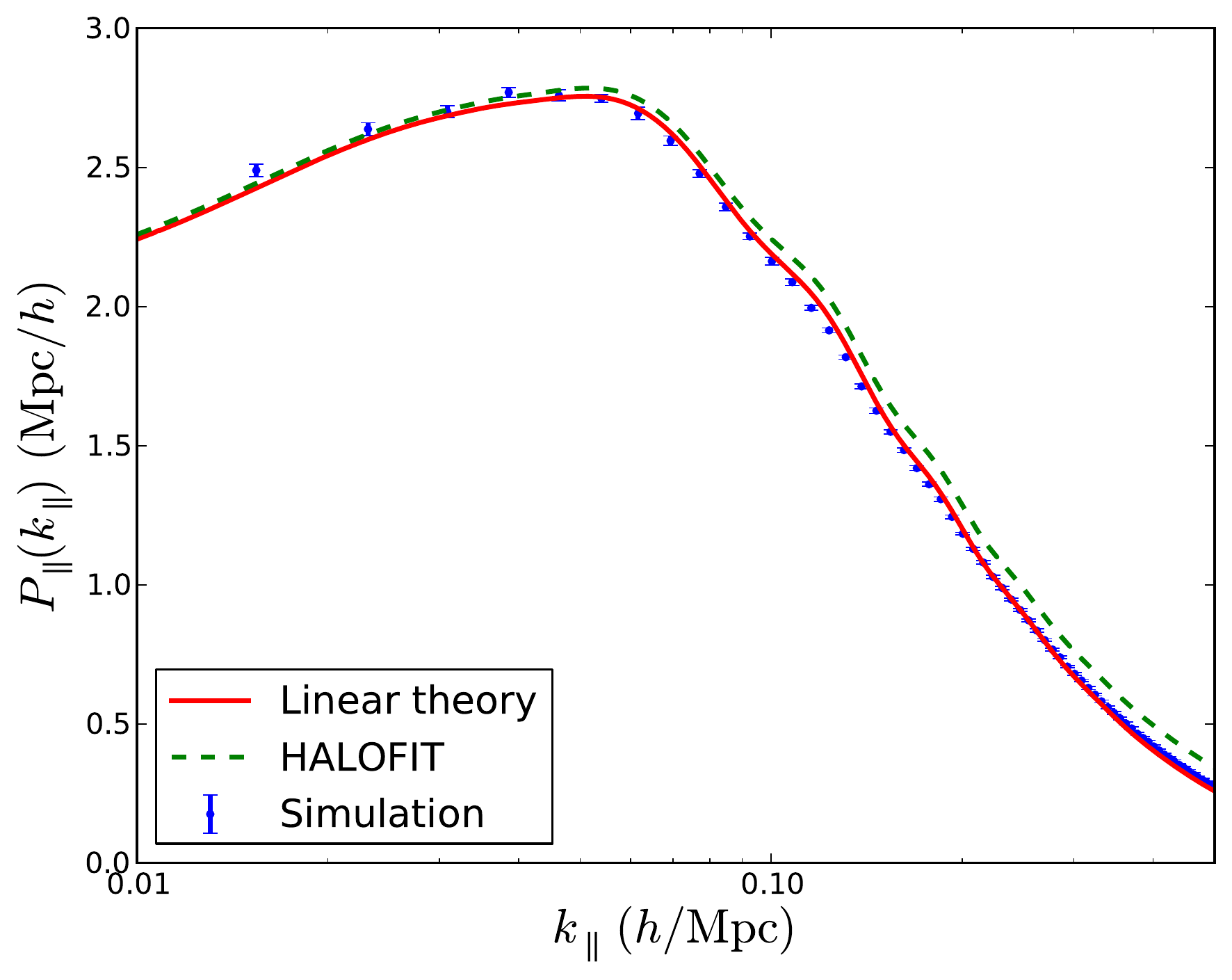}
        \caption{Cosmological radial power spectrum for a bin $1.4 < z < 2.1$, corresponding to
                 frequencies $457\, {\rm MHz} < \nu < 590\,{\rm MHz}$. The blue dots correspond
                 to the measurement from an \textsc{Accurate} simulation, while the solid red
                 line shows the linear theory prediction.}\label{fig:pkr_cosmo}
      \end{figure}
      \begin{figure*}
        \centering
        \includegraphics[width=0.90\textwidth]{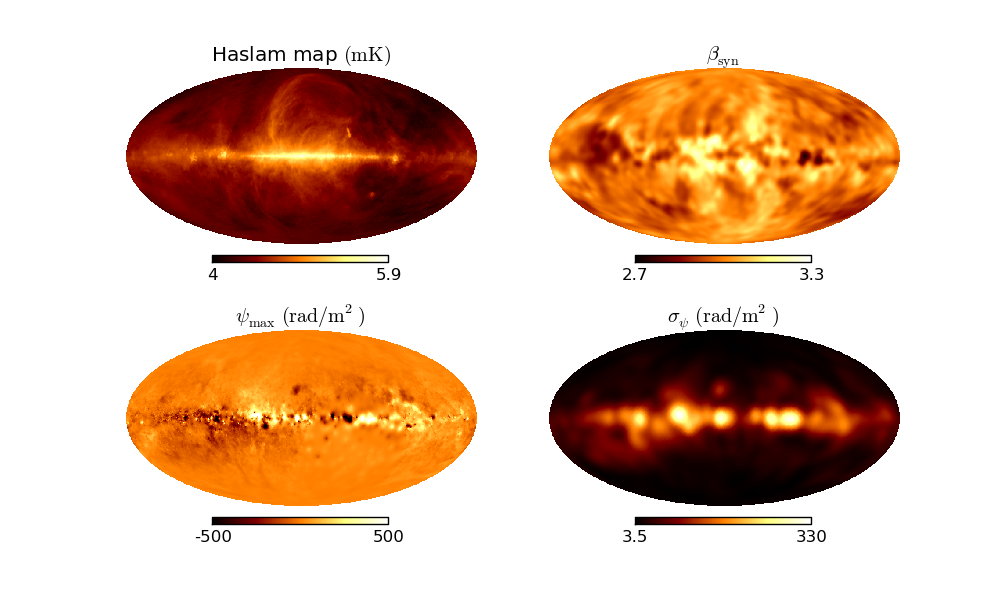}
        \caption{External datasets used to simulated the unpolarized and polarized synchrotron
                 foregrounds: the Haslam map (top left) (Temperature given in mK), the
                 synchrotron spectral index $\beta(\hat{\bf n})$ predicted by the Planck Sky
                 model (top right), the map of Faraday depths $\psi_{\infty}$ compiled by
                 \citet{2012A&A...542A..93O} (bottom left) and the Faraday widths
                 $\sigma(\hat{\bf n})$ estimated by smoothing $\psi_{\infty}^2$ (bottom right).}
                 \label{fig:data}
      \end{figure*}
      \begin{figure}
        \centering
        \includegraphics[width=0.45\textwidth]{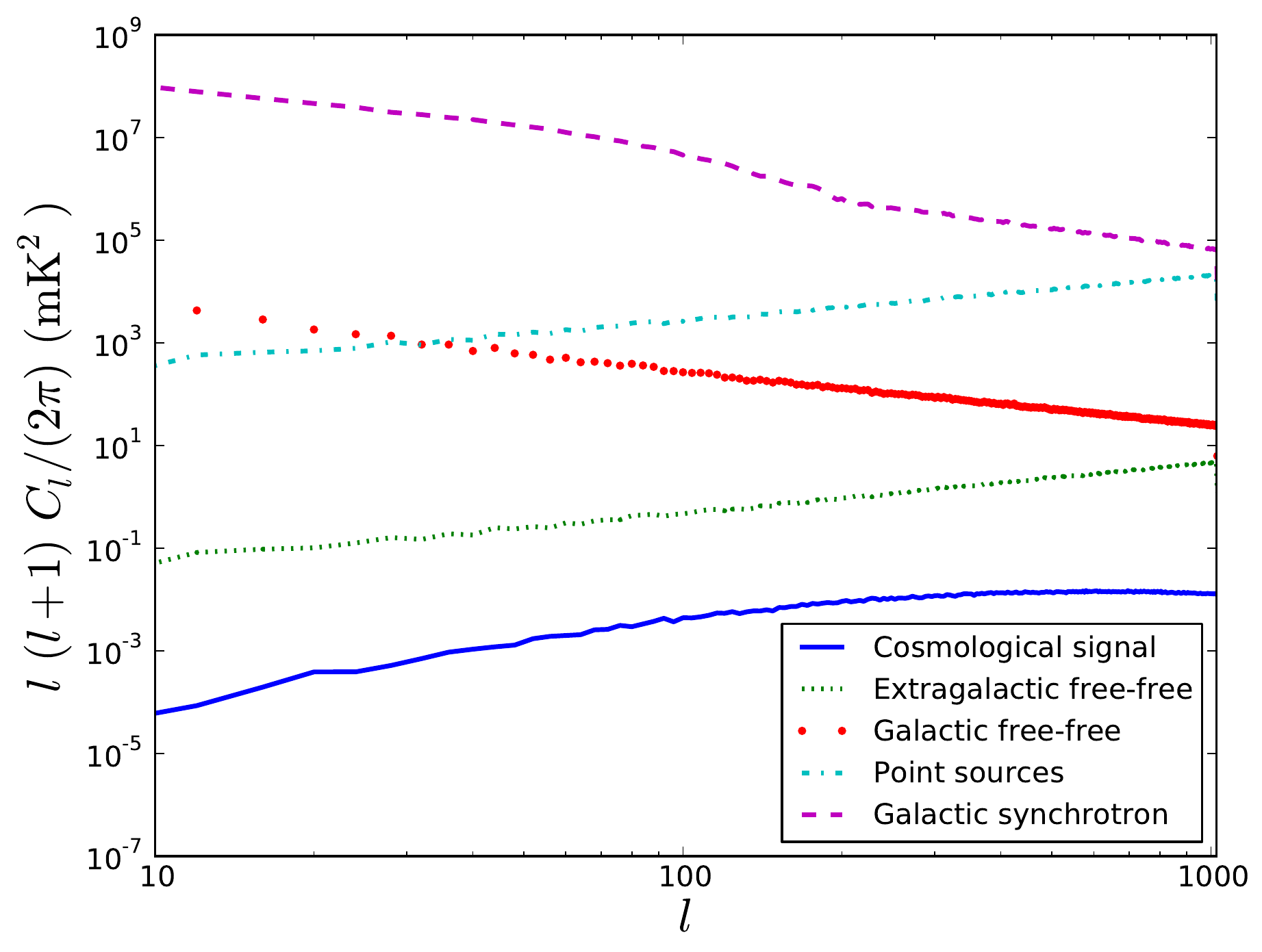}
        \caption{Angular power spectra of the different foreground components and the
                 cosmological signal at $\nu\sim550\,{\rm MHz}$.}\label{fig:cl_all}
      \end{figure}
      \begin{figure*}
        \centering
        \includegraphics[width=0.90\textwidth]{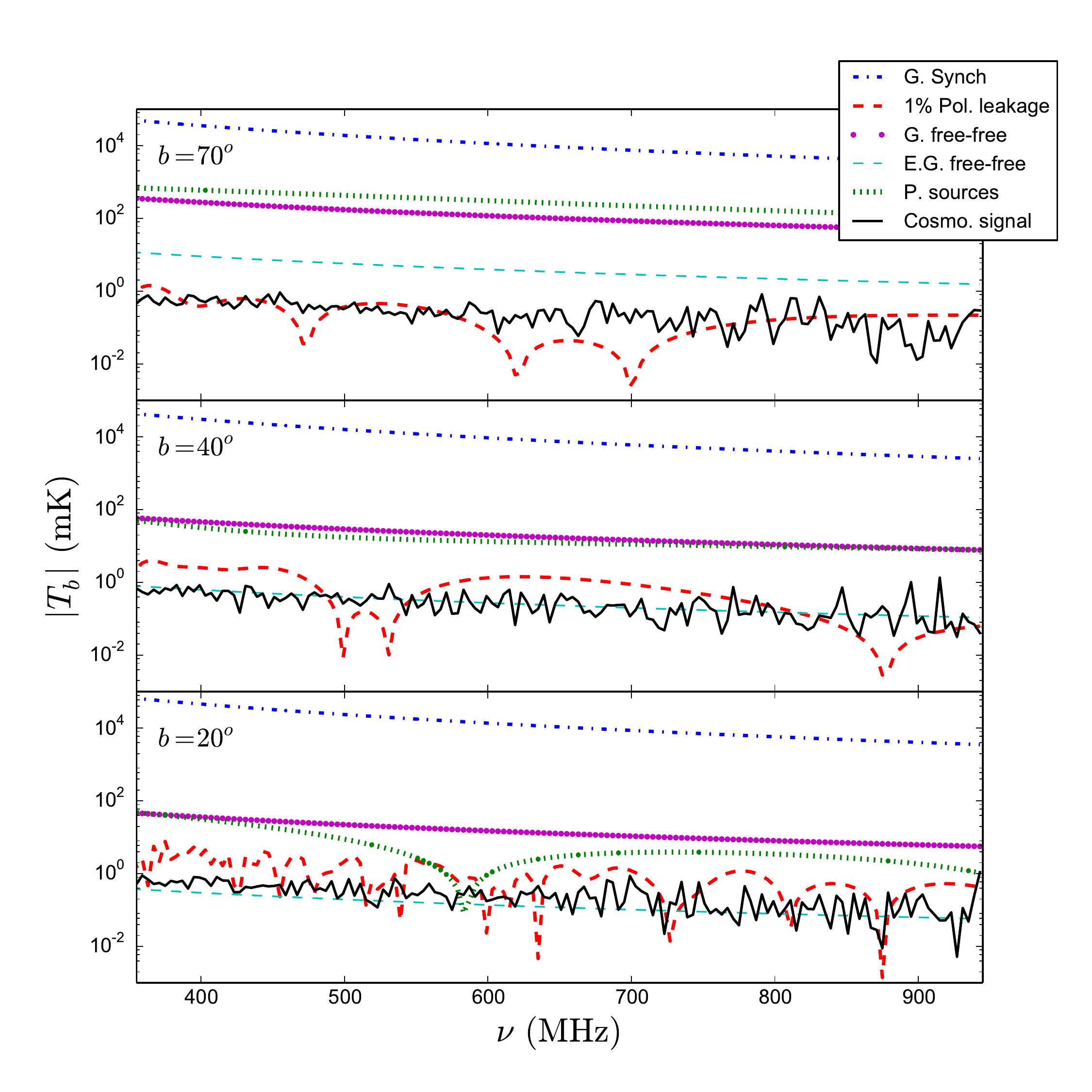}
        \caption{Frequency-dependence of the different foregrounds and the cosmological signal
                 along lines of sight with different galactic latitudes (given in the top
                 right corner of each panel). The effect of Faraday decorrelation increases as
                 we approach the galactic plane, making the subtraction of the polarization
                 leakage more challenging.}\label{fig:los_s}
      \end{figure*}
      For the cosmological signal we have used a $\Lambda$CDM model with parameters 
      $(\Omega_M,\Omega_b,\Omega_k,h,w_0,w_a,\sigma_8,n_s)=(0.3,0.049,0,0.67,-1,0,0.8,0.96)$,
      similar to the best-fit values found by \citet{2013arXiv1303.5076P}.
      The linear power spectrum used to generate the matter field was generated with {\tt CAMB}
      \citep{2000ApJ...538..473L} using these parameters. As a model for the neutral fraction we
      used $x_{\rm HI}(z)=0.008\,(1+z)$, which is a good fit to the two existing measurements at
      $z=0.01$ \citep{2005MNRAS.359L..30Z} and 1.5 \citep{2005ARA&A..43..861W}. For simplicity we
      assumed that the neutral hydrogen is unbiased with $b_{\rm HI}=1$.
      
      \subsubsection{Angular clustering}
        We have estimated the angular power spectrum of the cosmological signal for different
        frequency bins and compared it with the result predicted by linear theory. The
        comparison for the frequency bin at $\bar{\nu}=565\,{\rm MHz}$ ($z\sim1.5$) can be
        seen in Figure \ref{fig:cl_cosmo}. The theoretical power spectrum was corrected for the
        finite pixel size using a top-hat window function with a width equal to our {\tt HEALPix}
        pixel resolution (${\tt nside}=512\,\rightarrow\theta_{\rm TH}\sim0.06^o$). Rigorously
        speaking this correction would only be valid if all the pixels had a circular shape,
        however{, this correction is negligible compared with the window function due to the
        Gaussian beam, which removes most of the power above $l\sim400$. Figure \ref{fig:cl_cosmo}
        also shows an estimate of the non-linear angular power spectrum, using the 
        $P(k)$ predicted by {\tt HALOFIT} \citep{2003MNRAS.341.1311S}, however the
        difference with respect to  the linear prediction is negligible due to the
        large angular beam.

      \subsubsection{Radial clustering}
        We have also verified that our simulations reproduce the correct clustering pattern along
        the line of sight. We have done so by collecting all the frequency shells corresponding to
        the redshift interval $1.4<z<2.1$. For each pixel, this collection forms a vector with the
        values of the temperature fluctuation in different equispaced frequencies (which roughly
        correspond to equispaced radial distances). We perform a Fast Fourier Transform on each of
        these vectors to estimate $\delta_{\parallel}(k_{\parallel})$ and estimate the radial power
        spectrum by averaging over all the pixels.
        
        Figure \ref{fig:pkr_cosmo} shows the comparison between the thus measured radial power
        spectrum and the theoretical prediction outlined in section \ref{sec:th_clustering}
        using a radial window function corresponding to the comoving width of the spherical shells
        in our \textsc{Accurate} simulations $\delta\nu=0.7\,{\rm MHz}\rightarrow\delta
        r_{\parallel}\sim 4\,{\rm Mpc}/h$. As in the case of the angular power spectrum, the
        radial clustering is correctly reproduced by our simulations for a wide range of scales.
        The differences observed on large scales (small $k_{\parallel}$) can be understood as being
        due to the fact that, while the frequency width is fixed for all the shells, the
        corresponding physical widths are frequency-dependent and vary slightly ($\sim10\%$)
        across this frequency band, an effect that is not taken into account in our simplified
        model (section \ref{sssec:rad_cluster}). The non-linear prediction given by
        {\tt HALOFIT} is also shown as a green line, which deviates significantly from the linear
        theory model on small scales (large $k_\parallel$). We must caution the reader that this
        result is probably not quantitatively precise: the exact nature of different effects,
        such as redshift-space distortions or scale-dependent bias for HI is not well
        understood on non-linear scales, and therefore our simplified model is probably not
        realistic in this regime.

    \subsection{Foregrounds}
      As explained in section \ref{ssec:isotropic_foregrounds}, the isotropic foregrounds were
      generated simply as random Gaussian realizations of the $C_l$'s in Eq.
      (\ref{eq:clsck}), using the parameters in Table \ref{tab:clmodel}. On the other hand,
      simulating the polarized and unpolarized synchrotron foregrounds requires the use of 3
      additional external datasets: the $408\,{\rm MHz}$ Haslam map, the spectral index map
      from the PSM and the map of Faraday depths from \citet{2012A&A...542A..93O}.
      These datasets are displayed in Figure \ref{fig:data}.
      
      We have studied the angular distribution of these foregrounds as well as their frequency
      dependence. Figure \ref{fig:cl_all} shows the angular power spectra of the different
      foreground components compared to the cosmological signal that we expect to measure,
      which is several orders of magnitude smaller.
      
      Figure \ref{fig:los_s} shows the frequency dependence of the different components along
      lines of sight with different galactic latitudes ($b=70^o,\,40^o\,{\rm and}\,20^o$ from
      top to bottom). We can see that most foreground components are smooth in frequency, and
      should therefore be amenable to standard subtraction techniques. The leaked polarized
      synchrotron, on the other hand, has a non-trivial frequency structure, and could be
      extremely challenging to subtract. This problem becomes more important closer to the
      galactic plane, since the galaxy becomes ``thicker'' in Faraday-space, and the effects
      of Faraday rotation are more relevant. For this reason, special effort has been invested
      into verifying that our mock maps of the polarized synchrotron emission are statistically
      sensible.
      
      \subsubsection{Polarized foregrounds.}\label{sec:val_pol}
        \begin{figure}
          \centering
          \includegraphics[width=0.45\textwidth]{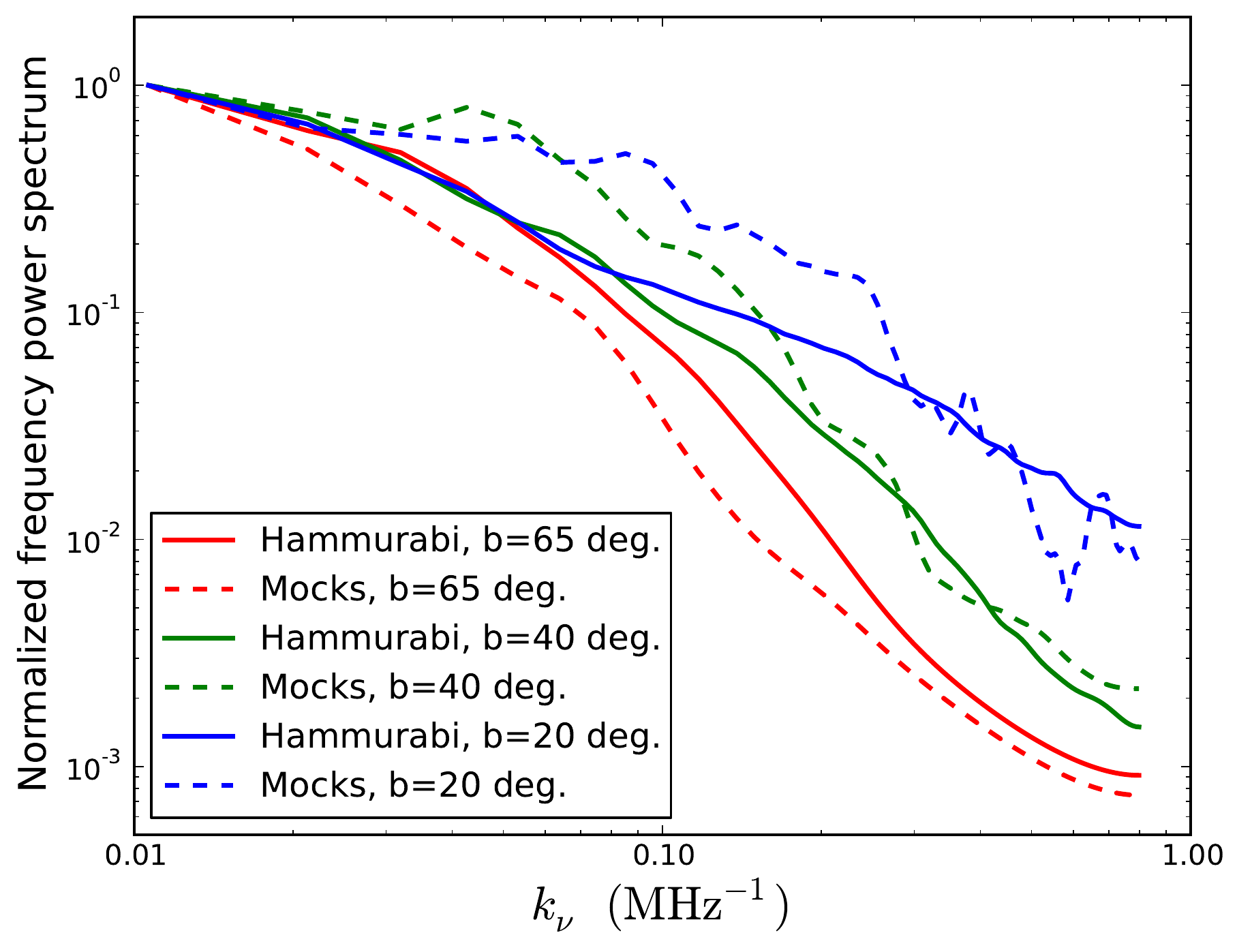}
          \caption{Frequency power spectrum at different galactic latitudes for {\tt Hammurabi}
                   (solid lines) and for the code presented in this paper (dashes). The curves
                   are normalized to 1 for the first non-zero wave number.}
                   \label{fig:noise}
        \end{figure}
        \begin{figure*}
          \centering
          \includegraphics[width=0.45\textwidth]{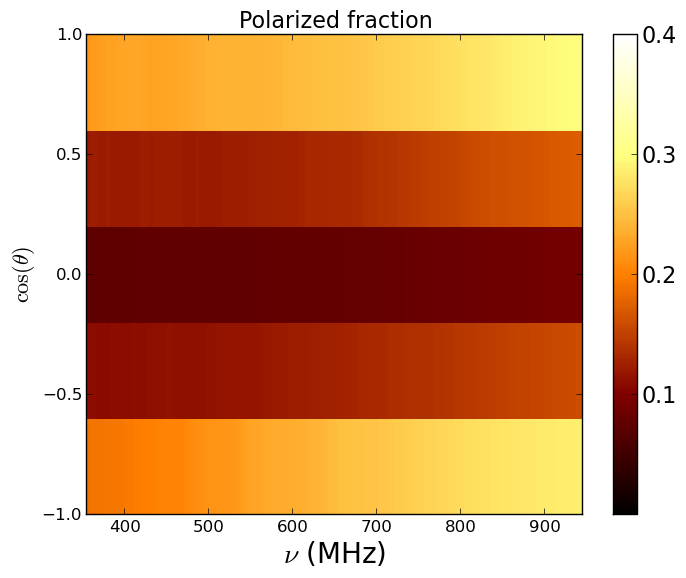}
          \includegraphics[width=0.45\textwidth]{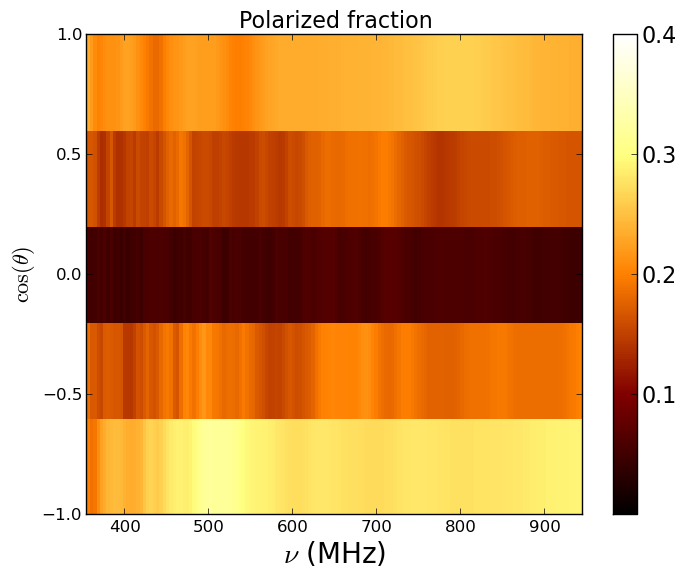}
          \caption{Polarized fraction on different galactic latitudes and frequencies
                   as reproduced by {\tt Hammurabi} (left panel) and our own code (right panel).}
                   \label{fig:polfrac}
        \end{figure*}
        According to the model described in section \ref{sssec:synch_pol}, the contribution of the
        polarized synchrotron to the foregrounds is described by two extra parameters, namely the
        Faraday-space correlation length $\xi_{\psi}$ and the polarization leakage fraction
        $\epsilon_p$. While the latter depends entirely on the instrument design, the former has a
        direct effect on the degree of frequency decorrelation, which is key to determining whether
        or not it will be possible to subtract it.
        
        In order to validate our model for the polarized synchrotron foregrounds we have run
        {\tt Hammurabi} using different parameters and spatial resolutions, and have compared the
        results with the maps generated by our code. In particular, we verified that the results
        shown below were qualitatively stable for the different models used by {\tt Hammurabi} to
        simulate the galactic magnetic field and cosmic ray electron density, the three-dimensional
        and angular resolution used by the code and the spectral tilt of the small-scale magnetic
        field. The fiducial simulation, for which we show results below, was run using the models
        for the magnetic field and CR density from \citet{2008A&A...477..573S}, three radial
        shells, an angular resolution parameter ${\tt nside}=256$ for the observation shell, a
        radial resolution of $\sim0.1\,{\rm kpc}$ and a Cartesian grid with resolution
        $\sim0.07\,{\rm kpc}$ (see \citet{2009A&A...495..697W} for an overview of these
        parameters). The rms variance of the small-scale magnetic field was set to $3\,
        \mu{\rm G}$. We generated synchrotron sky maps for 150 frequency bins between $945$
        and $355$ MHz.
        
        There are two main effects that we want our mock maps to reproduce:
        \begin{itemize}
          \item The degree of frequency decorrelation produced by the frequency-dependent Faraday
                rotation is the main source of complications in terms of foreground subtraction.
                Thus, we must make sure that this decorrelation is correctly reproduced by our
                model. In order to quantify this decorrelation we have computed the
                frequency-space power spectrum on different galactic latitudes for the
                {\tt Hammurabi} simulations and for our own mock datasets. This was done as
                follows:
                \begin{enumerate}
                  \item We take the emission in $Q$ for different frequencies for all the pixels
                        in a wide interval of galactic latitudes (we used $\Delta b=10^o$).
                  \item We Fourier-transform the emission from each pixel in frequency-space:
                    \begin{equation}
                      \widetilde{Q}(k_{\nu},{\bf n})\propto\int d\nu\,Q(\nu,{\bf n})\,
                      e^{i\,k_{\nu}\nu}.
                    \end{equation}
                  \item We compute the power spectrum $P(k_{\nu})\equiv
                        \langle|\widetilde{Q}(k_{\nu})|^2\rangle$
                        by averaging over all the pixels in the stripe.
            \end{enumerate}
            As mentioned above, Faraday rotation should be more important towards the galactic
            centre. Hence, frequency decorrelation should increase and $P(k_{\nu})$ should
            become bluer at low $b$. We can use this effect to constrain the only free
            parameter of our model: the Faraday-space correlation length $\xi_{\psi}$. This
            effect is explicitly shown in Figure \ref{fig:noise}, where we have plotted the
            normalized power-spectra for three latitude stripes at $b=65^o$, $40^o$ and
            $20^o$. We find that a value $\xi_{\psi}\sim0.5\,{\rm rad/m}^2$ is able to
            qualitatively reproduce this effect.
          \item The polarized fraction
            \begin{equation}
              \Pi\equiv \left\langle\frac{\sqrt{Q^2+U^2}}{I}\right\rangle,
            \end{equation}
            should decrease towards the galactic plane, since the Galaxy becomes ``thicker''
            and the incoherent emission from many regions cancels out. We have computed
            $\Pi$ for different frequencies in thick galactic latitude stripes both for the
            {\tt Hammurabi} simulations and for our maps. The result for the fiducial
            simulation is shown in Figure \ref{fig:polfrac}. As mentioned in section
            \ref{sssec:synch_pol}, this larger depolarization towards the galactic plane
            is automatically taken into account in our maps through the larger values of
            $\sigma_{\psi}$ computed from the map of \citet{2012A&A...542A..93O} (see
            the bottom right plot in Figure \ref{fig:data}).
        \end{itemize}
        
  \section{Conclusions}\label{sec:discuss}
    Intensity mapping of neutral hydrogen is potentially a very powerful observational tool
    to study the large-scale structure of the Universe at late times. In this paper we have
    presented a public code that can be used to generate simple and fast mock intensity
    mapping observations including both the cosmological signal and different galactic and
    extra-galactic foregrounds.
    
    We have verified that our methods are able to reproduce the most important features of these
    observations. In particular we have verified that the simulated cosmological signal follows
    the correct clustering statistics to second order both in the angular directions and along
    the line of sight. Furthermore, we have verified that our models for the different
    foregrounds mimic the most relevant observational effects, including their angular
    distribution and frequency decorrelation (particularly in the case of the polarized
    galactic synchrotron, which is probably the most troublesome foreground). Using our code, one
    full simulation, including cosmological signal and foregrounds with a spatial resolution of
    $\sim4\,{\rm Mpc}/h$ and a frequency bandwith of $\sim4\,{\rm MHz}$ can be generated in
    around 30 minutes using an 80-core machine - ideally we wish to speed this up by a factor of
    10 so as to be useful in Monte Carlo analysis. 
    
    We believe that these tools could be especially useful for a number of studies, such as:
    \begin{itemize}
      \item A thorough evaluation of different foreground-subtraction techniques.
      \item Optimizing the instrument design to suppress polarization leakage to the
        appropriate level.
      \item Assessing the viability of different cosmological analyses, such as BAO, RSDs, or
        primordial non-Gaussianity, in a realistic way (i.e. fully taking into account
        the effects of foregrounds subtraction).
      \item Studying the impact of different systematic effects on real observations.
      \item A complete statistical analysis of the uncertainties for different experiments.
    \end{itemize}
    
    As has been pointed out in the text, the methods used by our code have some shortcomings
    that we would like to address in future versions:
    \begin{itemize}
      \item We would like to implement alternative methods, such as higher order Lagrangian
            perturbation theory or quick particle-mesh algorithms, to generate the 3D matter
            density field. This would be necessary in order to explore higher-order moments
            of the clustering statistics.
      \item It would be desirable to have a more direct handle on the physical description
            of galactic synchrotron emission, which would require a 3D$+\nu$ line-of-sight
            integrator such as {\tt Hammurabi}.
      \item In its current version, our code only supports the simulation of $w$CDM models.
            In the future we plan to implement other non-standard cosmologies.
      \item The first public version of the code generates full-sky maps for both the signal
            and the foregrounds. We are currently working on a future version that will allow
            the user to simulate smaller patches of the sky and use the flat-sky approximation.
            This would also imply a major boost in speed and memory usage.
    \end{itemize}
    
    Our code is publicly available and can be found in
    {\tt http://intensitymapping.physics.ox.ac.uk/CRIME.html}. The code
    consists of a number of independent subroutines, written in C and Fortran 90, that generate
    the different components. Even though the code has been optimized to use a minimum amount of
    computer memory, intensity mapping provides naturally an immense amount of data (equivalent
    to one CMB map per frequency bin), and therefore these simulations can be expensive in terms
    of memory. For this reason some users may be more interested in obtaining the simulated maps
    used for this work rather than generating them. These maps can be found in the same URL.
    
  \section*{Acknowledgements}\label{sec:acknowledgements}
    We would like to thank: Tessa Baker, Gianni Bernardi, Philip Bull, Tzu-Ching Chang,
    Clive Dickinson, Matt Jarvis, Thibaut Louis, Sigurd N\ae{}ss and Jonathan Sievers for useful
    comments and discussions. The points raised by the journal referee also improved the quality
    of this work and we would like to acknowledge his contribution here. DA is supported by ERC
    grant 259505 and acknowledges the hospitality of the University of the Western Cape. PGF
    acknowledges support from Leverhulme, STFC, BIPAC and the Oxford Martin School. MGS
    acknowledges support from the National Research Foundation (NRF, South Africa), the South
    African Square Kilometre Array Project and FCT under grant PTDC/FIS-AST/2194/2012.

\setlength{\bibhang}{2.0em}
\setlength\labelwidth{0.0em}
\bibliography{paper}

  \appendix
  \section[]{Validity of the method}\label{app:nlin}
    As stated in section \ref{sec:intro}, in order to be able to generate fast mock
    observations it is often necessary to use approximate methods that reduce the
    computational complexity but that can lead to quantitatively incorrect results. This
    is the case of the method presented here, and therefore we would like to address
    the validity of some of these approximations here.
    
    \paragraph*{Non-linearities.}
      Even though small non-linear scales are not particularly relevant for intensity
      mapping, due to the low angular resolution of most experiments
      \citep{2014arXiv1405.1452B}, it is still interesting to analyze the regime in
      which the lognormal approximation described in section \ref{ssec:ln} yields valid
      results. We have done so by generating a realization of the density field in a
      comoving box at a fixed time (i.e. no lightcone evolution is applied). We have computed the
      power spectrum for this realization and compared it with the theoretical linear power
      spectrum provided by {\tt CAMB} and its non-linear prediction modelled by {\tt HALOFIT}
      \citep{2003MNRAS.341.1311S}.
    
      We have generated a lognormal realisation of the density field at redshift $z=0.7$ using
      the same cosmological parameters and spatial resolution that was used for the
      \textsc{Accurate} simulation (see table \ref{tab:sims}). Furthermore, in order to compare
      the validity of this method with the results for other techniques used in the literature
      \citep{2013MNRAS.428.1036M}, we have also generated the density field as predicted by
      first-order Lagrangian Perturbation Theory (1LPT - i.e. the Zel'dovich approximation
      \citep{1970A&A.....5...84Z}). For this we have generated the first-order Lagrangian
      displacements for a set of comoving tracers of the density field at the same redshifts
      (this is equivalent to setting the initial conditions for an N-body simulation at
      $z_{\rm init}=0.7$). We then computed the density field by cloud-in-cell interpolation
      of the positions of these tracers.
    
      Figure \ref{fig:pk_nlin} shows the ratio of the power spectra of these two realisations
      to the linear power spectrum (green circles for the lognormal field and red squares for
      1LPT), as well as the same ratio for the non-linear power spectrum (black line). It is
      evident from this figure that the lognormal transformation is not able to perfectly
      reproduce the non-linear power spectrum on all scales, although it does provide extra
      power with respect to the linear power spectrum for large $k$. The same is true for
      1LPT: shell-crossing at low redshift smooths the density field, and all power is lost
      on small scales. We can thus conclude that it is safe to use the method described in
      this work within the linear regime, which is the most relevant one for intensity mapping.
      However any analysis requiring a good description of the non-linear clustering
      should make use of a more sophisticated method.
      \begin{figure}
        \centering
        \includegraphics[width=0.45\textwidth]{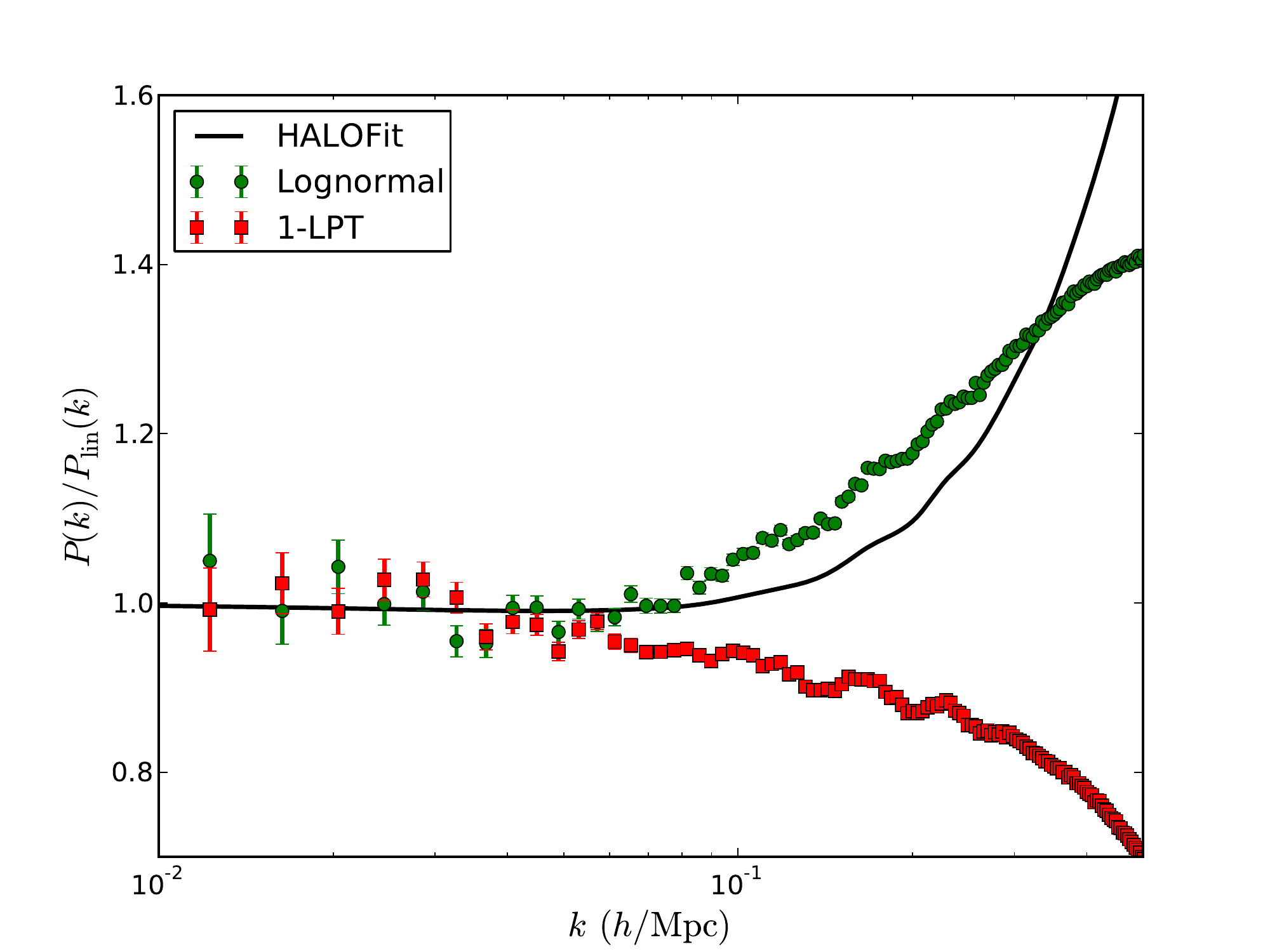}
        \caption{Ratio of the power spectrum computed from simulations to the linear power
                 spectrum. The simulated density fields were generated in at a fixed time
                 corresponding to $z=0.7$ in a box with the same spatial resolution that
                 was used for the \textsc{Accurate} simulation. We show results for a
                 lognormal density field (green circles), a 1-LPT field (red squares) and
                 the non-linear prediction given by {\tt HALOFIT} (black solid line).}
                 \label{fig:pk_nlin}
      \end{figure}
      \begin{figure}
        \centering
        \includegraphics[width=0.45\textwidth]{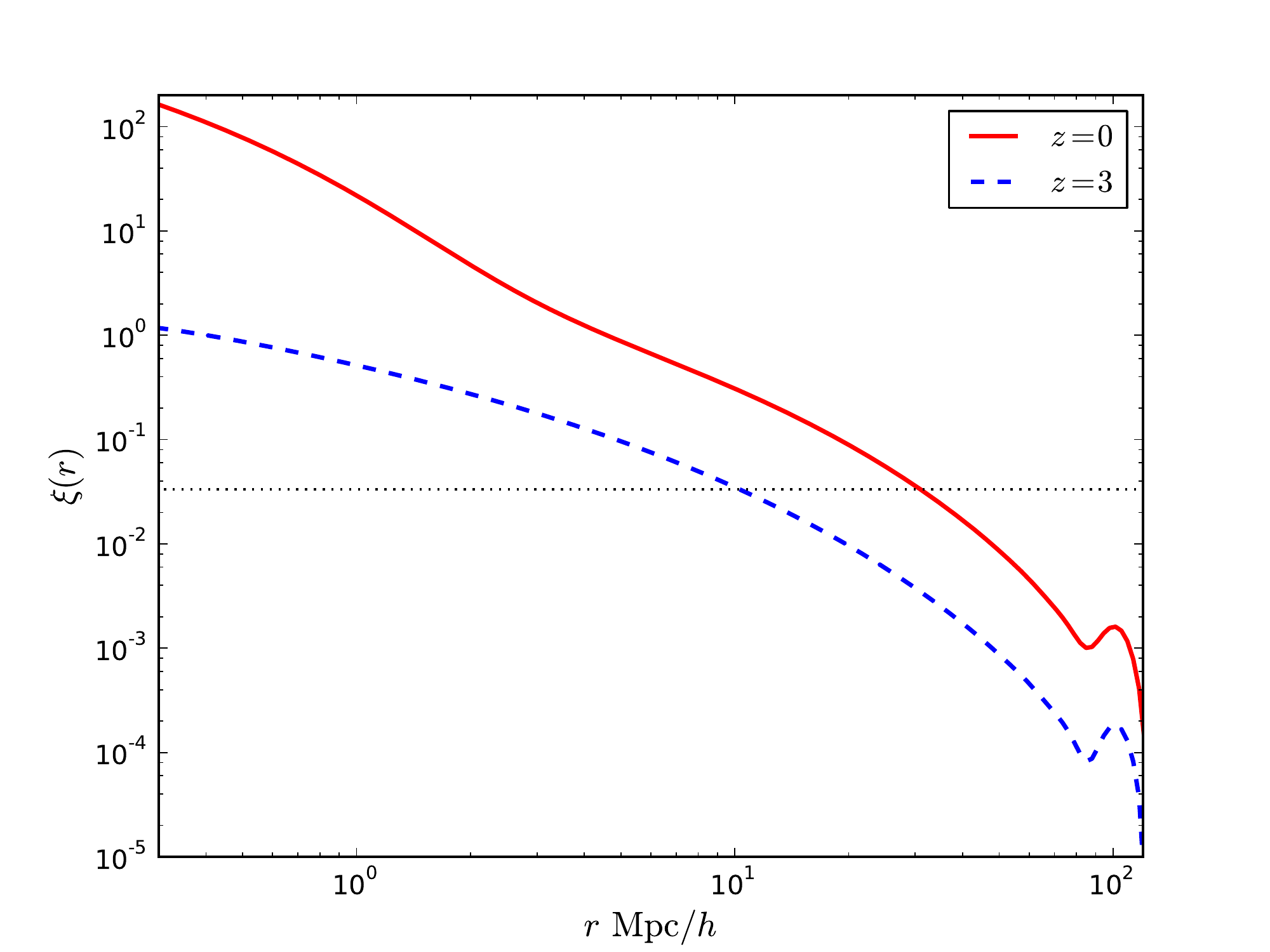}
        \caption{Non-linear two-point correlation function at $z=0$ (solid red) and
                 $z=3$ (dashed blue). The intercepts of these curves with the horizontal
                 dotted line show the approximate scale at which higher-order correlations are
                 comparable to $\xi(r)$.}
                 \label{fig:xi_highorder}
      \end{figure}
      
    \paragraph*{Scale-dependent bias.}
      Our code is able to include an evolving linear bias $b_{\rm HI}(z)$ that is constant
      for all scales. As is the case for most tracers of the density field, this can only be
      a good approximation for large scales, and it is important to know the regime in
      which scale-dependence becomes relevant. This has been studied in the literature,
      and in particular \citet{2012MNRAS.421.3570G} give constraints on both scale dependence
      and redshift evolution based on results from N-body simulations. Their results show
      that $b_{\rm HI}$ is constant in scale to a good approximation on scales
      $k\lesssim0.5\,h/{\rm Mpc}$ for $z<3$.
      
    \paragraph*{Higher-order correlations.}
      Since lognormal density fields are based on a local transformation of a Gaussian field,
      it is a well-documented fact that they are unable to reproduce the 3-, 4- and higher-order
      correlations produced by the non-linear gravitational collapse \citep{2014MNRAS.437.2594W}.
      This is not a problem for any science case that relies only on the two-point function,
      as is usually the case, however we can estimate the scale at which higher-order correlations
      become relevant.
      
      From the study of the three-point correlation function $\zeta$ in N-body simulations we
      know that it is possible to parametrize it as:
      \begin{equation}\nonumber
        \zeta(r_1,r_2,r_3)=Q_3\,\left[\xi(r_1)\xi(r_2)+\xi(r_2)\xi(r_3)+
                                      \xi(r_3)\xi(r_1)\right],
      \end{equation}
      where $Q_3$ is a function of the three relative distances $r_i$ that takes values
      $Q_3\sim\mathcal{O}(1-3)$ \citep{Peebles:1980,2005ApJ...632...29F}. We can thus get an
      idea of the scale at wich the amplitude of $\zeta$ is comparable to that of the
      two-point correlation $\xi$ by focusing on the equilateral configuration, $r_i=r$, in
      which case
      \begin{equation}
        \frac{\zeta(r,r,r)}{\xi(r)}\sim \mathcal{O}(1-3)\,3\,\xi(r).
      \end{equation}
      By choosing an arbitrary ratio of $\zeta/\xi=0.1$ we can then say that higher order
      correlations become important when $3\,\xi(r)\sim 0.1$. Between redshifts $z=0$ and 3
      this transition occurs in the range $r\sim30-10\,{\rm Mpc}/h$ (see figure
      \ref{fig:xi_highorder}).

\end{document}